# The TCP-modified Engset Model Revisited

Daniel Zaragoza[*] – February 2014

**Abstract.**
*We revisit the TCP-modified Engset model proposed by Heyman et al. in [1]. The model deals with the superposition of a limited number of TCP connections alternating between file transmission and silence in a web-like fashion. We consider homogeneous sources only. (a) We take into account the effects of slow start and limited receiver window as well as small average file sizes. (b) We propose an alternative way for calculating the average connection rate in the superposition. (c) From the model we propose a way for calculating the queuing behavior; i.e., the overflow probability. (d) From this last point, we propose a new link buffer sizing rule. Comparison with extensive simulations shows that the average rate and duration, as well as, link utilization are accurately predicted for exponentially distributed file sizes. For longer tail distributions, the model remains accurate provided the receiver window is adjusted appropriately. The accuracy increases with increasing load. As concerns the queuing behavior, the same observation applies. Finally, the revisited model cannot be used to predict losses larger than about 1%. The model overestimates loss rates above that threshold.*

***Keywords**. TCP modeling, short connections, superposition, slow start, limited receiver window.*

***Contents**.*

# 1   Introduction

In the present work, we revisit the TCP-modified Engset (TCP-Engset for short) model proposed by Heyman, Lakshman, and Neidhardt in [1]. The model deals with the performance of a superposition of Transmission Control Protocol (TCP) connections alternating between file transmission and silence periods in a web-like or ON-OFF fashion. File sizes and OFF periods are random variables. Connections share a common multiplexing link. The number of connections simultaneously active is limited to a maximum. The main idea of the model is that competing TCP connections share the bottleneck capacity equally in times of congestion.

[*] daniel.zaragoza@laposte.net

Connections are homogeneous meaning that all parameters are the same for all sender-receiver pairs: file size distribution, round trip time, ingress and egress links, packet size, etc. Throughout the report the OFF periods are exponentially distributed with an average of 1 second.

The contributions of the present work are:

(i)     We take into account the effects of slow start and limited receiver window in the calculation of the single connection rate and ON time. We also take into account small average file sizes.

(ii)    We propose an alternative way of calculating the average connection rate. Extensive simulations show that taking the average of the two ways give more accurate results than either one alone.

(iii)   We propose a simple way of determining the queuing behavior.

(iv)    We propose a new link buffer setting rule based on the revisited model.

The report is organized as follows. The next section discusses related work. In section 3 we give definitions, which include a TCP background, a discussion of the file size distributions we use here, and a discussion of the network settings and simulations setup. In section 4 we introduce the revisited TCP-Engset model. The section includes an overview of the approach, the calculation of the ON time for a single connection taking into account the effects of slow start and limited receiver window, the calculation of the state probabilities of the model, and the calculation of the queue statistics when a single connection uses a link. Section 5 is devoted to the comparison of model and simulations results. We first compare model and simulation results for exponentially distributed file sizes for a wide range of loads. We then investigate the effects of varying size distribution and receiver window on the model results and state probabilities. Finally, we evaluate the performance when packet drops occur due to a limited buffer at the multiplexing link. We first propose and evaluate a buffer setting rule. We then turn to the analysis of the effects of packets drops under high load, then by varying the buffer size. In section 6 we provide a commented model summary and our conclusions.

In the remainder of the report, we use the following abbreviations and notation.

Acknowledgement(s) is abbreviated as ack(s).

Duplicate acknowledgement is abbreviated as DA (we do not use delayed acknowledgements).

Triple duplicate acknowledgement is abbreviated as TD. A TD triggers a fast retransmit.

A successful fast recovery is abbreviated as FR.

Timeout is abbreviated as TO.

Congestion avoidance is abbreviated as CA.

Round trip time is abbreviated as RTT.

We use the word 'flight' to mean a group of contiguous packets.

Bandwidth delay product is abbreviated as BDP.

$W_{CA}$, window at which the sender switches from slow start to congestion avoidance.

$W_{FR}$, sender window just after FR occurred and "normal" operation resumes in CA.

$W_R$, receiver window, a.k.a. $rwnd$.

$S$, slow start threshold, a.k.a. $ssthresh$.

$\theta$, amount of data outstanding at the sender.

$W$, sender window.

$\sigma = W - \theta$, amount to send.

$\lfloor x \rfloor$, is the integer part of $x$.

$L_{SS}$, last packet sent in slow start when $W_R$ is reached for the first time.

$L$, loss rate. Measured as $L = \frac{\#Dropped}{\#Appearing}$

$F$, average file size in packets and $P$ packet size in bytes.

$N$, maximum number of active connections, and $\langle n \rangle$ the average number of these active connections.

$h$, average rate per connection and $ON = F \times P/h$ the average duration.

## 2 Related work

We discuss here related work on the modeling of short TCP connections.

The most directly related work is the paper by Heyman, Lakshman, and Neidhardt [1], where the TCP-Engset model has been first proposed. Their equation (25) is at the heart of the model. In words, the model assumes perfect capacity sharing between active sources whenever their number is larger than the ratio of the output to input link capacities. The authors derive a buffer sizing rule for window-synchronized TCPs in congestion avoidance under different drop policies so that the link is fully utilized; see their equation (22). For tail-drop policy, we call this rule the BDP-rule. They consider extensions of the model with multiple classes of users having the same access link capacity. They also discuss insensitivity properties.

Our starting point is the capacity sharing idea and the state probabilities calculation from [1]. The differences of the present work with [1] are as follows (See section 4.1 and Table 11 for details). We consider only homogeneous sources. First, we consider small or very small average file size and include the effects of slow start and limited receiver window. Second, we propose an alternative way of calculating the average connection rate. It turns out that taking the average of the two ways give more accurate results. Third, we propose a way to evaluate the queuing behavior which leads us to a buffer sizing rule that is different from the BDP-rule.

Less directly related works are summarized by Bonald in chapter 2 of [2]; see also the included references. The model could be called TCP-modified Erlang as connections arrive according to a Poisson process and congestion/blocking is replaced by capacity/resource sharing. We do not discuss here the relationship between the TCP-Engset and TCP-Erlang models.

To conclude this section, we note that the number of TCP variants in operation on the Internet has greatly increased between the measurements performed by Medina *et al.* [3] in 2005 and those performed by Yang *et al.* [4] in 2011 (their figure 1 presents a nice synthetic overview of TCP congestion controls). From Rewaskar *et al.* [5] we learn that implementations in the popular operating systems Linux and Windows have a $minRTO$ of 200 ms instead of the 1 second specified in the standard (see next section). The 'vagaries' of real world transfers using TCP are analyzed by Qian *et al.* [6]. Overall, the papers mentioned here tend to indicate that the job of modeling the performance of TCP will be a tough one.

## 3 Definitions

We provide here background on TCP, the file size distributions we use, the network settings, and the simulation setups.

### 3.1 TCP background

In this section we develop the aspects of TCP that are the most relevant to our purpose. (i) TCP is a window-based transport protocol. (ii) It provides flow-control. (iii) It provides congestion control via the slow start and congestion avoidance algorithms. (iv) Retransmission timeouts are important for its performance. Finally, (v) the sending of new data is triggered by the arrival of acks within the timeout limit.

The relevance to our purpose of modeling the performance of 'short' transfers is as follows. First, when no packets are dropped, a connection proceeds in slow start until either the transfer is complete or the receiver window is reached. In slow start without delayed acks, a TCP sends two packets for each ack received, which means that two packets are sent for one leaving the network. When the receiver window is reached the correspondence is one for one. When a capacity mismatch exists between input(s) and output(s) at a multiplexing node the slow start algorithm puts some 'stress' on the buffer of the node. This 'stress' also depends on the RTT of the connection, the smaller the RTT the greater the 'stress', see section 4.4. On the other hand, the smaller the RTT the faster a connection completes. Second, when packets can be dropped – here, only due to buffer overflow –, a connection is particularly 'fragile' at the beginning and at the end where a TO can occur. Finally, the arrival of data packets (at the receiver) and acks (at the sender) need not be regular for the proper operation of TCP; therefore, TCP traffic is said to be 'elastic'.

RFC4614, [7], provides a roadmap to TCP-related documents, as of 2006. Documents of interest here that update previous ones cited in RFC4614 are: RFC5681, [8], – TCP congestion control –, RFC6582, [9] – TCP NewReno –, and RFC6298, [10] – TCP RTO calculation.

TCP is first specified in RFC793, [11]; further details are clarified in RFC1122, [12].

TCP is a window-based protocol that provides a reliable, byte-oriented, data transfer between processes which, most commonly, run on different computers (hosts) themselves located on different networks. In the original specification, TCP can send and receive variable-length "segments". However, most implementations refrain from sending variable length segments unless some conditions are met, e.g., sending a segment empties the send buffer or the Nagle's algorithm is disabled [13].

TCP provides reliability "*by assigning a sequence number to each octet transmitted, and requiring a positive acknowledgment (ACK) from the receiving TCP. If the ACK is not received within a timeout interval, the data is retransmitted.*" Further, "*The sequence number of the first octet of data in a segment is transmitted with that segment and is called the segment sequence number. Segments also carry an acknowledgment number which is the sequence number of the next expected data octet of transmissions in the reverse direction.*" How the timeout is calculated is specified in RFC6298 and is discussed below. In the present work, we use a TCP that is packet-oriented; that is the basic unit of transmission is the packet rather that the octet.

TCP is a window-based protocol which provides flow control to avoid a sender overwhelming a receiver. For that purpose, "*The receiving TCP reports a "window" to the sending TCP. This window specifies the number of octets, starting with the acknowledgment number, that the receiving TCP is currently prepared to receive.*" At the sender and at any time, the amount of data outstanding and not yet acknowledged – a.k.a. the "flight size" – is $\theta = nxt - una$, where $nxt$, is the sequence number of the next unit to send and $una$ is the sequence number of the still unacknowledged unit. Under no circumstances and at no time $\theta > W_R$, where $W_R$ is the receiver window, a.k.a. $rwnd$. $W_R$ is not necessarily fixed and varies according to the speed at which the application "reads" the received data during the course of a connection. Note that $una$ changes only with advancing acks, while $nxt$ increases when sending new data and temporarily decreases back to $una$ when retransmitting missing data.

A form of indirect/implicit flow control may also exist at the sender. RFC793 states: "*When the TCP transmits a segment containing data, it puts a copy on a retransmission queue and starts a timer; when the acknowledgment for that data is received, the segment is deleted from the queue. If the acknowledgment is not received before the timer runs out, the segment is retransmitted.*" In practice, implementations [13], maintain a "send buffer" (a.k.a. the socket buffer) which contains the unacknowledged data and which is fed with data from the user buffer; see also Figure 2 below. The size of this buffer may have impact on the TCP performance. In the present work we do not deal with this matter and assume ideal operation at the sender.

TCP also provides congestion control (RFC5681) by way of the slow-start and congestion avoidance algorithms. For that purpose two new variables are used: the congestion window, $cwnd$ and the slow-start threshold, $ssthresh$, noted $S$ here. A TCP switches between slow-start and congestion avoidance according to the relative values of $cwnd$ and $S$. More precisely, "*The slow start algorithm is used when cwnd < ssthresh, while the congestion avoidance algorithm is used when cwnd > ssthresh. When cwnd and ssthresh are equal, the sender may use either slow start or congestion avoidance.*" Further, "*The initial value of ssthresh SHOULD be set arbitrarily high (e.g., to the size of the largest possible advertised window), but ssthresh MUST be reduced in response to congestion.*" Here, we use congestion avoidance when $cwnd > S$ meaning that $W_{CA} = S + 1$. Rephrasing RFC5681, during slow-start $cwnd$ is increased by one for each increasing ack received; during congestion avoidance $cwnd$ is increased by one for each $cwnd$ of increasing acks received. To improve performance when the receiver acknowledges every other segment and to discourage misbehaving receivers, a TCP may use Appropriate Byte Counting as per RFC3465, [14]. In our TCP, the receiver acknowledges every segment received and the sender counts the increasing arriving acks before increasing the congestion window in congestion avoidance; this is to avoid numerical rounding errors.

The sending of data (new or retransmission) by TCP is governed by the following equations.

$$W = \min(cwnd, W_R), \theta = nxt - una, \text{ and } \sigma = W - \theta. \tag{1}$$

Where $\theta$ is the amount outstanding ("FlightSize") when the sending TCP function (e.g., `tcp_output()`) is called, $\sigma$ is the amount to send, starting from $nxt$. When a group of packets are to be sent, i.e., when $\sigma > 1$, these are sent within a software loop at 'code speed' [13] with no network-dependant spacing between packets. Although proposals such as TCP-pacing have been made, these studies remain theoretical with no active specification or commercial implementation to our knowledge. This mode of operation has implications on the burstiness of the traffic produced by a TCP. These implications are well-known and recommendations in RFC6582 are made to limit the possible bursts of packets sent at once, for example, using an additional variable $maxburst$. In the present report, $W, \theta$ and $\sigma$ are in unit of packets (segments). Further, as $W_R$ never changes, we set $cwnd \leq W_R$ $^{(*)}$. No attempt is made to limit the data bursts.

TCP uses two methods to detect missing data: retransmission timeout and duplicate acks.

When the retransmission timer fires, a timeout occurs and $cwnd = 1$, $nxt = una$, thus $\sigma = 1$ and the missing packet is retransmitted. If this is the first retransmission of the packet, $S = \min(\lfloor\theta/2\rfloor, 2)$. $S$ is left unchanged in case of multiple retransmissions of the same packet. The connection proceeds in slow-start mode. As a consequence, duplicate packets can be sent after a timeout; duplicates can also be lost, though.

The second method is via duplicate acks, typically three, which means that four consecutive acks request the same sequence number. From RFC6582, "*After receiving 3 duplicate ACKs, TCP performs a retransmission of what appears to be the missing segment, without waiting for the retransmission timer to expire.*" This is called fast retransmit. We abbreviate this event as TD, for Triple Duplicate acks. "*After the fast retransmit algorithm sends what appears to be the missing segment, the "fast recovery" algorithm governs the transmission of new data until a non-duplicate ACK arrives.*" We abbreviate fast recovery as FR. When a TD occurs, $S = \min(\lfloor\theta/2\rfloor, 2)$ (same as for timeouts), and the missing packet is retransmitted. The previous value of $nxt$ is restored to allow the sending of new data with further arriving DAs and $cwnd = S + 3$. Note that no new data is sent here even if it were possible (specifically, when $\theta = 4$). For each subsequent duplicate ack received, $cwnd += 1$ and new data may be sent. Finally, "*When the next ACK arrives that acknowledges previously unacknowledged data, a TCP MUST set cwnd to ssthresh (the value set in step 2). This is termed "deflating" the window.*" At this point, depending on the TCP version used, FR occurs or not, more details are given below. Assume FR occurs; thus, $cwnd = S$. "Normal" operation resumes, that is, $cwnd$ is increased either to $W = S + 1/S$ if $W_{CA} = S$ or to $W = S + 1$ if $W_{CA} = S + 1$ and new data is sent depending on $\sigma = W - \theta$ being positive. In both cases, the connection proceeds in congestion avoidance mode.

---

$^{*}$ It is not advisable to do so when $W_R$ can change during the course of a connection.

If any one of the retransmissions is also lost then FR does not occur, instead, a timeout follows. We do not address this matter here.

For TCP Reno, FR occurs on receipt of the first ack requesting unacknowledged data. The loss of only one packet in the previous window of data has no further consequence. If there were $d$ drops, then another TD or a timeout follows. Under the assumption $W_{CA} = S + 1$, let $W$ be the window at which drops occur, let $\delta$ be the distance, in packets, between the first two drops, if $\delta < W - S - d$ then a timeout follows; otherwise, a TD follows if, in turn, enough acks return from the new packets sent during the TD-FR period.

For TCP NewReno (RFC6582), FR occurs when all data outstanding at the time TD occurred is acknowledged. Note that the loss of the new data sent between TD and FR lead to a new congestion event. Because NewReno is used in the present report, we give an example of operation in Figure 1.

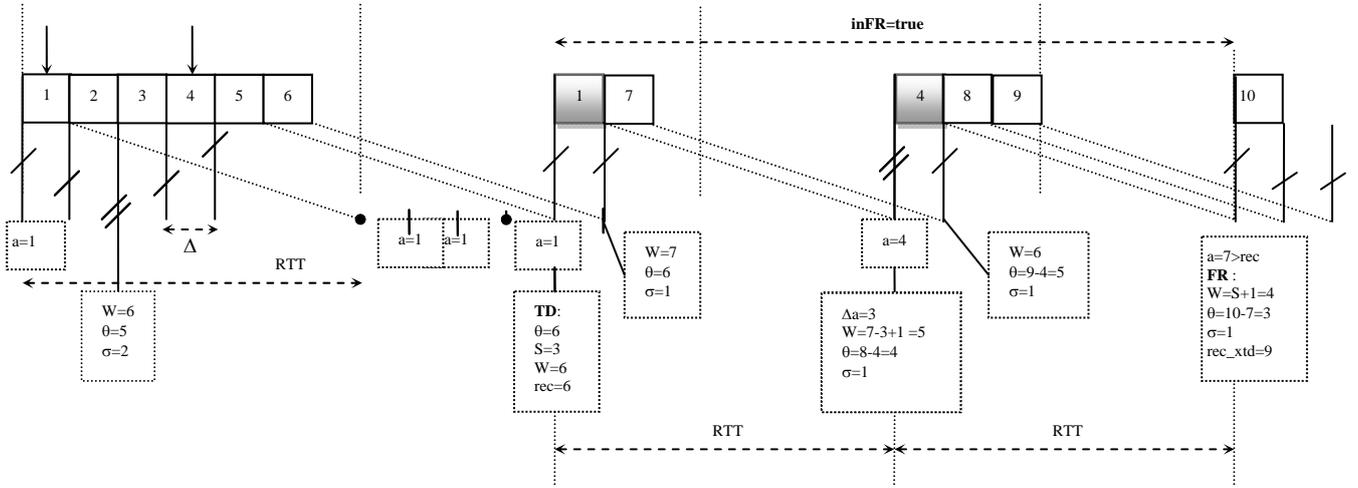

**Figure 1. Example of TD-FR with TCP NewReno from the sender perspective.**

Without loss of generality for the present purpose, we renumber the first loss as '1' and resynchronize time. When TD occurs (on the arrival of the ack from '5'), the missing packet is retransmitted, $recover = 6$, and both $S$ and $cwnd$ are adjusted. $W_R$ is large enough and does not change so that $W = cwnd$. The flag $inFR = true$. On receipt of the subsequent ack (from '6'), new data is sent. When the first loss is recovered, FR is not yet over for NewReno. Following RFC6582, the missing packet is retransmitted and a new data packet is sent. Further new data is sent on receipt of the ack from '7', which was new also. FR is over on the receipt of the acknowledgement $a = 7 > recover$; $inFR = false$, $cwnd = S + 1 < W_R$; normal operation resumes and new data is sent. In the same RTT, the receipt of the ack from '8' will lead to $cwnd += 1/\ cwnd$ as it must in congestion avoidance. For the purpose of collecting further statistical data on TOs and TDs, we record the highest sequence number sent during the TD-FR period in the variable $rec\_xtd$; if one or more of these new packets is lost a TD or a timeout will follow FR.

For the 'patient' version, the retransmit timer is set only on TD; thus, up to $\left\lfloor \frac{RTO}{RTT} \right\rfloor$ drops can be recovered. In the present work, we use the 'impatient' version which sets the retransmission timer on each retransmission in fast recovery. More details are given is section 5.4.

As concerns timeouts, RFC6582 states the following: "*After a retransmit timeout, record the highest sequence number transmitted in the variable recover, and exit the fast recovery procedure if applicable.*" Subsequent TD events are ignored if they do not cover more than $recover$. This is to avoid spurious TDs.

Our TCP follows RFC6582 only for the first retransmission after a timeout. For subsequent retransmissions of the same packet $recover = 0$, thus all TD events are enabled. Further, a $TDoff$ flag is used for the purpose of collecting timeout statistics.

The final point of this section is the calculation of the retransmission timeout, which depends on the RTT of the connection, as defined in RFC6298. There are four points to consider: (i) Setting of the initial RTO (ii) Computing a proper RTO. (iii) RTT measurement. (iv) Setting the RTO for multiple retransmissions of the same packet.

(i). The initial RTO, when no RTT measurement has been made yet, is set to 1 second (it was previously 3 seconds). The "backing off" discussed below applies.

(ii). When RTT measurements, $R$, say, are available, the RTO is computed in the following two steps; assuming precise timing (zero clock granularity).

    (a)    First measurement:
$$SRTT \leftarrow R, \; RTTVAR \leftarrow R/2, \; RTO \leftarrow \max(1 \text{ s}, \; SRTT + 4 \times RTTVAR).$$
    (b)    Subsequent measurements:
$$RTTVAR \leftarrow (1 - 1/4) \times RTTVAR + 1/4 \times |SRTT - R|,$$
$$SRTT \leftarrow (1 - 1/8) \times SRTT + 1/8 \times R,$$
$$RTO \leftarrow \max(1 \text{ s}, \; SRTT + 4 \times RTTVAR).$$

With $R$ constant, the first measurement gives $RTO \leftarrow \max(1 \text{ s}, \; 3 \times R)$. On the tenth measurement, $RTO \leftarrow \max(1 \text{ s}, \; 1.113 \times R)$, and on the twentieth, $RTO \leftarrow \max(1 \text{ s}, \; 1.006 \times R)$, etc. Therefore, for a large connection, if $R > 1$ second $RTO = R$. A number of popular implementations use a $minRTO$ of 200 ms instead of 1 second for the setting of the RTO value.

(iii). RTT measurements are performed, at least, once per RTT outside of any loss event, and using new data. This means that retransmissions are not timed and measurements are not performed during TDFR periods. On FR, normal operation resumes and the first segment sent is timed. Generally speaking, RTT measurements are taken once per "flight" at the beginning of the flight. An implication is that spurious timeouts can occur with too large (in relative terms) $W_R$ and router buffers over slow links. An example is given in section 4.2.

It is important not to confuse timed segments for the purpose of measuring the RTT and setting the retransmission timer on segments. Under normal operation, on retransmission after a timeout, and on FR, the timer must be set on each packet sent (timed or not) – in practice, when a burst of packets is sent, the timer is set on the first of the group. During a TDFR period, the timer can be set either on the first retransmission only (patient version) or for each retransmission (impatient version).

(iv). Backing off the RTO. When a packet is retransmitted the RTO is multiplied by two. When the same packet is retransmitted multiple times the RTO is multiplied again by two, up to four times ($2^5 = 64$) or up to the maximum RTO of 60 seconds. According to Windows implementation public documents, the retransmission of the same packet is limited to five times, at which point the connection is aborted. Our TCP does not limit the number of retransmissions but $maxRTO$ is 64 seconds.

Our TCP follows the above cited RFCs with the following specifics. The receiver acknowledges every packet received. The send buffer is "infinite". The initial window is 2 packets. The RTO value and RTT measurements are set with precise timing; that is, the $G$ of RFC6298 is zero. The retransmit timer is set for each retransmission during fast recovery. Our TCP does not send new data on the first two duplicate acks before TD occurs (no "limited transmit"). We use $W_{CA} = S + 1$.

Further, connection setup and termination are not simulated. Unlike a real TCP, both sender and receiver know how many packets are to be sent and received when the connection starts. For each connection, the initial sequence number is '1'; there is no possibility of arrival of packets from previous connections. At the sender, a connection is over when the last packet is acknowledged. At the receiver, a connection terminates when all packets are received in sequence. The other aspects of the protocol such as a changing receiver window, restart after idle, etc. are not simulated.

In order to be complete, we conclude this section with a simplified but typical software and buffer architecture inspired from [13] shown in Figure 2. The purpose is to insist on the fact that buffer settings and software efficiency are of importance for TCP/IP performance.

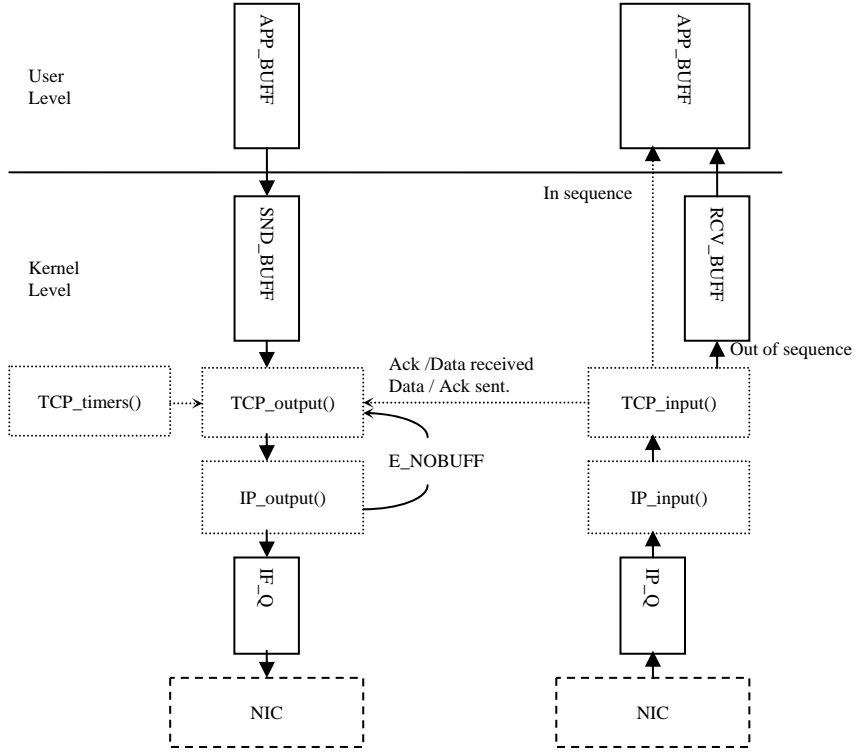

**Figure 2. Simplified buffer and software implementation inspired from [13].**

### 3.2 File size distributions

For the selection of file sizes we use the following distributions: exponential (E), exponential-body-Pareto-tail (EP), and exponential-body-exponential-tail (EE). Files have a minimum size also.

For a continuous random variable $X$ we note $G(x) = \mathbb{P}(X > x)$.

For an exponentially distributed variable with a minimum, $A$, we have,

$$G(x) = \begin{cases} 1 & , \ x < A \\ e^{-(x-A)/B} & , \ x \geq A \end{cases}.$$

The average value is $\bar{X} = A + B$.

For a Pareto distribution we have

$$G(x) = \begin{cases} 1 & , \ x < k \\ \left(\frac{k}{x}\right)^{\alpha} & , \ x \geq k \end{cases}.$$

The average value is $\bar{X} = k\frac{\alpha}{\alpha - 1}$. The average is finite when $\alpha > 1$. The variance is finite when $\alpha > 2$. Note that the minimum is $k$. If $\alpha$ and the minimum are given, then the average is fixed. This does not allow for much flexibility.

In order to have more flexibility, we use a distribution with an exponential body and a Pareto tail.

$$G(x) = \begin{cases} 1 & , & x < A \\ e^{-(x-A)/B} & , & A \leq x < k \\ H \times \left(\frac{k}{x}\right)^{\alpha} & , & x \geq k \end{cases}, \text{ with } H = e^{-(k-A)/B}.$$

$H$ gives the proportion of file sizes that follow the Pareto distribution. We set $H = 10\%$, meaning that 10% of the files can be large to very large. Note that this distribution is continuous but not derivable.

The average is $\bar{X} = A + \frac{Hk}{\alpha - 1} + B(1 - H)$.

We fix the minimum file size $A$, typically $A = 3$, $\alpha = 1.5$, and $H = 10\%$.

For a given $\bar{X}$, $B = \frac{\bar{X} - A\left(1 + \frac{H}{\alpha - 1}\right)}{1 - H - \frac{Hln(H)}{\alpha - 1}}$. Once $B$ is calculated, $k = A - Bln(H)$.

In order to further check that the simulation results we obtain are not due solely to the Pareto distribution, we use a distribution with an exponential body and an exponential tail. We also set $H = 10\%$, meaning that 10% of the files can be large (as defined below by the parameter $C$) but not extremely large because of the exponential decay.

$$G(x) = \begin{cases} 1 & , & x < A \\ e^{-(x-A)/B} & , & A \leq x < k \\ He^{-(x-k)/C} & , & x \geq k \end{cases}, \text{ with } H = e^{-(k-A)/B}.$$

The average is $\bar{X} = A + B(1 - H) + CH$.

With the condition $B > 0$, we have that $C < C_{max} = \frac{\bar{X} - A}{H}$. We fix $C$ via $C = \omega\bar{X} < C_{max}$, thus, $\omega < \frac{\bar{X} - A}{H\bar{X}}$.

With $C$ now given (as well as $\bar{X}$, $A$, and $H$), $B = \frac{\bar{X} - A - CH}{1 - H}$ and $k = A - Bln(H)$.

For $F = 5$ packets we use $\omega = 3.9$, while for $F \geq 12$ $\omega = 7$, that gives $B > 0$ for $\bar{X} > 10$.

The following Table 1 gives the measured proportion of file sizes per intervals. The intervals are chosen to correspond to 'flights' when the receiver window is 44 packets. The second column gives the sender window span during slow start in the interval. The values are the proportions for the E and EP distributions; the average size per interval is also given in parenthesis. The calculated values are very close to the measurements.

**Table 1. Measured proportion of file sizes per size intervals.**

| | | F=5, 800kC | | F=12, 800kC | | F=22, 450kC | | F=36, 250kC | | F=80, 120kC | | F=120, 80kC | |
|---|---|---|---|---|---|---|---|---|---|---|---|---|---|
| [f1-f2] | W | Expo | EP k=5.4 | Expo | EP k=17.2 | Expo | EP k=34.1 | Expo | EP k=57.8 | Expo | EP k=132.3 | Expo | EP k=200 |
| 3-6 | 3-4 | 82.7 (4.2) | 92.8 (3.7) | 32.2 (4.6) | 43.6 (4.6) | 16.8 (4.7) | 22.9 (4.6) | 10.0 (4.7) | 13.7 (4.7) | 4.4 | 6.1 | 3 | 4 |
| 7-14 | 5-8 | 17.0 (8.4) | 5.05 (9.2) | 39.9 (9.9) | 41.2 (9.7) | 28.5 (10.2) | 34.5 (10.1) | 19.4 (10.3) | 24.7 (10.3) | 9.5 | 12.5 | 6.2 | 8.8 |
| 15-30 | 9-16 | 0.3 (16.4) | 1.45 (20.1) | 23.2 (20.3) | 11.1 (19.1) | 19.1 (42.2) | 29.6 (21.0) | 27.1 (21.9) | 30.2 (21.6) | 16.2 | 20.2 | 11.7 | 14.8 |
| 31-62 | 17-32 | | 0.46 (41.6) | 4.6 (38.8) | 2.7 (41.9) | 4.2 (79.3) | 9.0 (40.6) | 26.9 (43.9) | 22.6 (42.5) | 23.6 | 26.5 | 18.9 | 22.5 |
| 63-128 | 33-44 | | 0.05 (175) | 0.13 (71.0) | 0.93 (86.3) | 0.14 (147) | 2.7 (85.9) | 14.3 (85.2) | 5.8 (85.9) | 26.5 (91) | 24.0 (89) | 26.1 (92.5) | 26.8 (91) |
| 129-256 | 44 | | 0.017 (355) | | 0.32 (174) | | 0.88 (176) | 2.2 (159) | 1.91 (175) | 16.2 (176) | 7.1 (173) | 23.0 (180) | 16.2 (172) |
| 257-512 | 44 | | | | 0.11 (352) | | 0.31 (352) | 0.05 (292) | 0.71 (346) | 3.6 (325) | 2.4 (348) | 9.9 (341) | 4.4 (348) |
| 513-1024 | 44 | | | | 0.04 | | 0.11 | | 0.25 | 0.14 (590) | 0.85 (697) | 1.3 (620) | 1.6 (693) |
| 1025-2048 | 44 | | | | 0.01 | | 0.04 | | 0.08 | | 0.3 | 0.03 | 0.58 |
| 2049-4096 | 44 | | | | | | 0.015 | | 0.03 | | 0.1 | | 0.19 |
| > 4096 | 44 | | | | | | | | 0.02 | | 0.05 | | 0.11 |

The average file size above $H$ can be calculated by

$$F_H = k_H + \int_{k_H}^{\infty} G(x)\, dx,$$

where $k_H$ is calculated from $H = e^{-(k-A)/B}$

To conclude this section, Figure 3 illustrates $G(x)$ for $F = 12$ packets for the different distributions we use here. The EP distribution is fairly smooth while the EE one is fairly abrupt. Note also that the average size above $H$ is 33, 52 and 89 packets for the E, EP, and EE(7) distributions respectively. Above 400 packets, the EP distribution is above the EE(7) one, yet the average above $H$ is larger for EE(7).

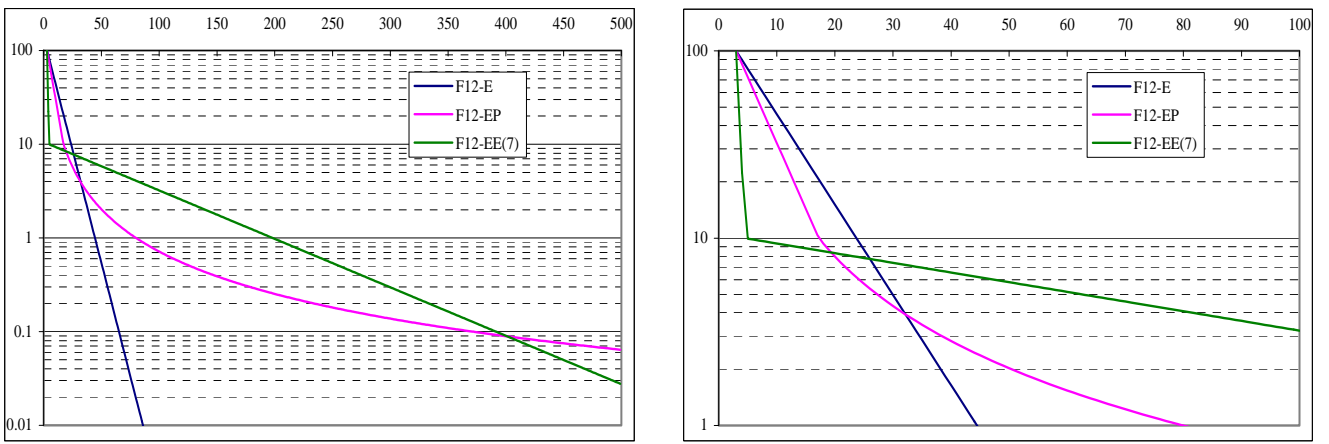

**Figure 3. File size distributions for F = 12 packets.**

### 3.3 Network settings

The network setting used in this report is illustrated in Figure 4. There are $N$ TCP senders to $N$ TCP receivers. Each receiver has its own receiving (egress) link $L_3$ of capacity $C_3$, each of these links has a buffer such that no packets are ever dropped due to buffer overflow. Each sender has its own sending (ingress) link $L_1$ of capacity $C_1 = 100$Mbps. The router multiplexing these inputs has a buffer of size $B$ packets. Depending on $B$, packets may be dropped due to buffer overflow. The link $L_2$ between the ingress and egress routers has a capacity $C_2 = C$. Note that $N$ is not necessarily the number of hosts; however, the maximum number of simultaneously active TCP connections is limited to $N$.

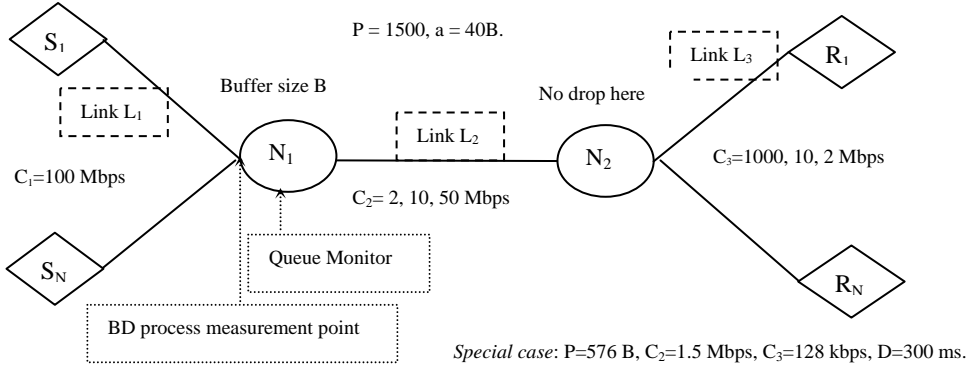

**Figure 4. Network setting.**

With reference to Figure 4, we note $\Delta_1 = P/C_1$, $\Delta_2 = P/C$, $\Delta_3 = P/C_3$ the transmission times of data packets on each link. The round-trip time for a single packet – not counting queuing – is $RTT_0 = D + (P + a) \times (\frac{1}{C_1} + \frac{1}{C} + \frac{1}{C_3})$, where $D$ is the – constant – round-trip propagation delay. $RTT_0$ is the time required to send a packet and receive its acknowledgement on an empty path. By linearity, the RTT is the same from any point of view; sender output, node input, etc.

We let $\beta = RTT/\Delta$. $\lfloor\beta\rfloor$ gives the number of packets that can 'fit' in a generic link of capacity $C$. $\beta$ is similar to the well-known $BDP = (\frac{D}{C})/P$.

Let us introduce now a sufficiently long TCP connection with receiver window limited to $W_R$.

We let $r = \frac{\lfloor\beta\rfloor}{W_R}$. $r \geq 1$ means that, in the long term, the link (path) is not saturated; $r < 1$ means that the link (path) is saturated.

When the link is saturated, the long-term queue contents is $Q = W_R - \lfloor\beta\rfloor + 1$. The '1' accounts for the packet in transmission. The number of packets outstanding is $W_R$, $\lfloor\beta\rfloor$ are 'in the link', thus the difference is in the waiting queue and one in transmission. When queuing occurs, we have $RTT = RTT_0 + Q \times \Delta$.

With reference to Figure 4, either links $L_2$ or $L_3$ can be saturated. $C_3 = 1$ Gbps means that there is no limitation whatsoever to reach the receiver after $L_2$.

Table 2 gives saturation data for different simulation scenarios (with $\Delta_1 = P/C_1 = 120$ μs fixed). $C$ stands for $C_2$ or $C_3$.

**Table 2. Saturation data for different simulation scenarios.**

| C (Mbps) | Δ (ms) | RTT (ms) | β (pkts) | W_R (pkts) | r | W_R / β (%) | Link Condition |
|---|---|---|---|---|---|---|---|
| 2 | 6 | 50 | 8.3 | 8 | 1.00 | 100 | Not saturated (but at onset) |
| | | | | 12 | 0.67 | 150 | Saturated, Q=5 |
| | | | | 20 | 0.40 | 250 | Saturated, Q=13 |
| | | | | 44 | 0.18 | 550 | Saturated, Q=37 |
| | | 100 | 16.7 | 8 | 2.00 | 50 | Not saturated |
| | | | | 12 | 1.33 | 75 | Not saturated |
| | | | | 20 | 0.80 | 125 | Saturated, Q=5 |
| | | | | 44 | 0.36 | 275 | Saturated, Q=29 |
| 10 | 1.2 | 50 | 41.7 | 8 | 5.13 | 19.5 | Not saturated |
| | | | | 12 | 3.42 | 29.3 | Not saturated |
| | | | | 20 | 2.05 | 48.8 | Not saturated |
| | | | | 44 | 0.93 | 107 | Saturated, Q=4 |
| | | 100 | 83.3 | 8 | 10.38 | 10 | Not saturated |
| | | | | 12 | 6.92 | 14.5 | Not saturated |
| | | | | 20 | 4.15 | 24 | Not saturated |
| | | | | 44 | 1.89 | 53 | Not saturated |

### 3.4 Simulation setup

The approaches and models developed here are compared with packet-level simulations. With reference to Figure 4, at the entry of node $N_1$ we place the 'Birth and Death' (BD) measurement module in order to compare simulation and theoretical results, as well as obtain statistical data. Node $N_1$ also includes a 'Queue Monitor' module to obtain statistical data about the queue process. The data collected include the distribution of the queue contents measured as a time average, the queue contents on packet arrival, as well as other statistical data.

Node $N_2$ behaves like a switch with a buffer for each outgoing link; setting this buffer to $W_R$ guarantees that no packets are ever dropped on the output links.

There are $N$ senders pair wise linked to $N$ receivers. Both senders and receivers know the size of the file to be transferred. The sender waits for an exponentially distributed OFF time with an average of 1 second (5 seconds in the special setting) between connections. Sources and sinks are homogeneous meaning that distributions, packet size, RTT, and receiver windows are the same for a given simulation.

The simulation runs until $N_c$ connections have been completed. $N_c$ depends on the average file size, $F$, with $F = 5, 12, 22, 36, 80,$ and 120 packets. $N_c$ is chosen such that there are about 10 millions packets ($F \times N_c$) to send. For $F = 5$ and F = 12 , $N_c = 800k$, for = 22 $N_c = 450k$, for $F = 36$ $N_c = 250k$, for $F = 80$ $N_c = 120k$, for $F = 120$ $N_c = 80k$ connections. Links $L_1$ have always the same capacity $C_1 = 100$Mbps. For a given simulation, links $L_3$ have also the same capacity. We verified that, given the large number of connection, results differ very little from simulation to simulation with the same settings; we therefore present the results for one simulation only.

For each of the $F$ indicated above we run simulations for the triplets (2M, 1G, R50-R100-R300), (10M, 2M, R50-R100-R300) and (10M, 1G, R50-R100-R300). The first element indicates the capacity of link $L_2$, the second that of links $L_3$, and the third the RTT in ms; other simulations follow the same notation. We also run simulations for the special case (1.5M, 128k, D300); that is, $C_2 = 1.5$Mbps, $C_3 = 128k$bps, and $D = 300$ms. This setting corresponds to the one used in [1] with the difference that it is now the receivers instead of the senders that have a 128kbps link. The second difference is that the window is limited to $W_R$.

Finally, to collect further drop statistics, TCP packets headers are augmented with information such as whether the packet was sent in slow start ($W \leq S$) or congestion avoidance ($W > S$), whether the packet is a retransmission after TO, a retransmission in fast recovery (a TO will then follow), etc. Adding such information is clearly possible in a simulation but not in a real implementation.

# 4   TCP-Engset revisited

In this section, we develop our approach to the TCP-Engset model. We first provide an overview. We then turn to details. Finally, we evaluate the queuing statistics for a single connection.

## 4.1   Overview

The method we use to model the performance of ON-OFF file transfers with TCP follows the following steps. In the following, we let $C$ be the capacity of the multiplexing link.

For a given network scenario and average file size $F$, we first calculate:

1- The average ON time for a connection alone on the path,

$$ON_0 = mRTT_0 + n\Delta^*, \text{ with } \Delta^* = \max (\Delta_i). \tag{2}$$

The calculation is detailed in section 4.2. We then calculate the connection peak rate

$$h_0 = \frac{F \times P}{ON_0}. \tag{3}$$

This gives

$$s = \left\lfloor \frac{C}{h_0} \right\rfloor. \tag{4}$$

With the knowledge of the average $OFF$ time, we can compute the – open loop – traffic load/demand on the link of capacity $C$ with $N$ sources; that is,

$$\rho_0 = \langle n \rangle \times h_0 / C = \frac{ON_0}{ON_0 + OFF} \times N \times h_0 / C. \tag{5}$$

The average number of active sources is $\langle n \rangle = \sum_1^N kP_k = a \times N$, with $a = \frac{ON}{ON+OFF}$. We allow $\rho_0 > 1$.

2- With the additional knowledge of the number of sources $N$, we calculate the $P_j(s)$, which make up the core of the TCP-Engset model. The calculation is detailed in section 4.3.

3- From the $P_j(s)$, we calculate the per connection rate and ON time in two different ways.

Let $P_{OL} = \sum_{s+1}^N P_k$, $n_{UL} = \sum_1^s kP_k$, and $n_{OL} = \sum_{s+1}^N kP_k$.

From equation (27) and (28) of [1] we have,

$$\text{Throughput} = T = h_0 \times n_{UL} + C \times P_{OL} \text{ and } \rho_{(1)} = T/C. \tag{6}$$

$$h_{(1)} = T/\langle n \rangle. \tag{7}$$

An alternative way to calculate $h$ is,

$$h_{(2)} = \frac{C}{v_{OL}}, \text{ with } v_{OL} = \frac{n_{OL}}{P_{OL}}. \tag{8}$$

We also have $ON = \frac{F \times P}{h_{(2)}}$, $a = \frac{ON}{ON + OFF}$, and finally

$$\rho_{(2)} = a \times N \times h_{(2)} / C \leq 1, \tag{9}$$

the – closed loop – traffic load/demand.

A detailed in section 5.1, $h_{(1)}$ is a little too high and $h_{(2)}$ a little too low; therefore, we use

$$h = \frac{h_{(1)} + h_{(2)}}{2}, ON = \frac{F \times P}{h}, \text{ and } \rho = \frac{\rho_{(1)} + \rho_{(2)}}{2}. \tag{10}$$

NOTES: (a) Equation 8 makes sense only for $N > s$. (b) In high overload, $P_{OL} \approx 1$ and both approaches give the same connection rate. (c) We have $\rho \leq 1$; when $P_{OL} \approx 1$ we have $\langle n \rangle = v_{OL} = C/h$, which gives $\rho = 1$. (d) It turns out that $\rho$ is the link utilization whether packets are dropped or not.

The second part, which is about the queue contents, is made of the following three steps. It is important to note that because of the assumption of instantaneous capacity sharing the standard technique used for open-loop ON-OFF sources cannot be applied because there are no overload states. Therefore, new approaches are needed. Our very simple model is one of them. The idea is that the ON time gets extended due to queuing.

Whenever a link on the path can be saturated, we renormalize $ON_0$ to $RTT_0$; that is, we recalculate the $m$ and $n$ of eq. (2) so that $n < \min\left(\lfloor \beta_i \rfloor\right)$.

4- We calculate

$$RTT = \frac{ON - n\Delta^*}{m}; \text{ but we also have } RTT = RTT_0 + Q\Delta, \text{ which gives us}$$

$$Q = \frac{RTT - RTT_0}{\Delta}, \tag{11}$$

the average queue contents. Note that $\Delta$ refers to the multiplexing link.

5- Assuming an exponential decay of the queue contents, we have $Q = \rho \times \eta$, which gives

$$\eta = Q / \rho. \tag{12}$$

6- Finally, we start from $G(x) = \mathbb{P}(Q > x)$ = proportion of time $Q > x$. We let $G(x)$ = proportion of arrivals that see $Q > x$; simulations show that this is a good approximation. For a finite buffer of size $B$, we approximate the loss ratio – the proportion of arrivals that see a full buffer– as

$$L = \rho e^{-\left(B/\eta\right)}.$$

This gives

$$B = \eta \text{Ln}(\rho/L). \tag{13}$$

A detailed analysis of the effects of packet drops is given is section 5.4.

Figure 5 illustrates the first two steps of the approach. First, a conversion to a 'classical' open-loop ON-OFF source is performed. Second, assuming instantaneous and perfect capacity sharing when $j > s$ connections are active, each connection adapts its sending rate to the situation. Clearly, the approach is rate-based and does not take into account the burstiness of the window-based TCP sending patterns both in slow start and congestion avoidance. In the first part, it assumes that no packets are dropped. In the second part, when losses occur, retransmissions are not accounted for.

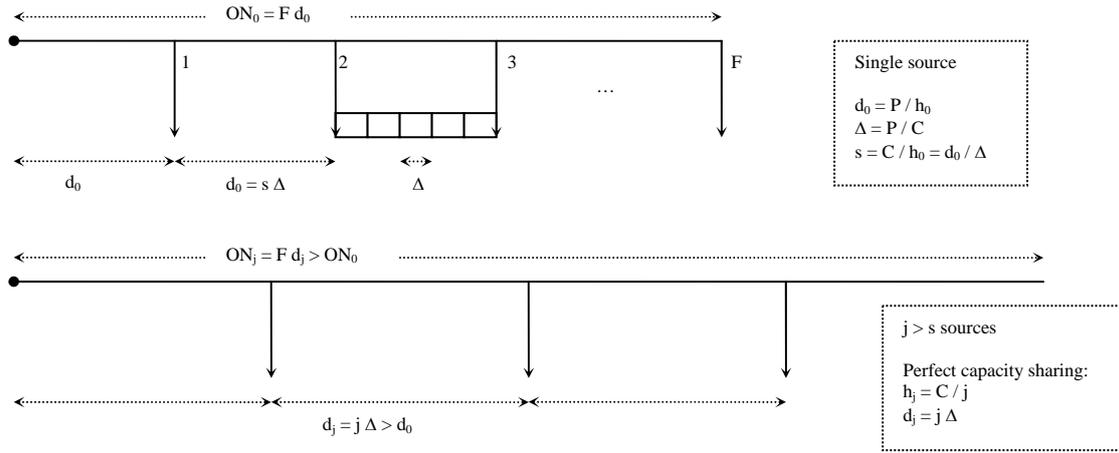

**Figure 5. Illustration of the TCP-Engset approach.**

### 4.2   Calculation of $ON_0$ and $h_0$

There is no simple and handy formula for calculating the ON time when a single connection of size $F$ uses a given path; we calculate it case by case. The receiver acknowledges every packet and has a limited window of $W_R$ packets. We let $\Delta^* = \max{(\Delta_i)}$ and $\beta^* = \min{(\beta_i)}$. The following Figure 6 illustrates the case for $F = 5$ packets. We use the word 'flight' to mean a group of contiguous packets. When the link is not saturated flights are separated. When the link is saturated at some window, the flight persists until the end of the transfer. In both the analysis and the simulator, when simultaneous arrival and departure occur, priority is given to departure.

The ON time is taken as the time elapsed between the sending of the first packet and the reception of the ack for the last packet, which means that the connection is now over. If we used the time of the sending of the last packet instead – noted $ON_{SR}$ – the OFF time would not be independent of the ON time due to the fact that this last packet may be delayed or even dropped (a timeout would then follow). We have also experimented with measurements of $ON_{SR}$; to avoid $ON_{SR} = 0$ with only one packet to send we have set the minimum file size to three packets. We keep this minimum value even if we do not consider further the use of $ON_{SR}$ in the present work.

NOTE: The BD process measurement module measures a ON time that is $P/C_1$ shorter than the one defined above for the calculation. This is because the process is updated only on the arrival of the first packet of the connection.

The following Table 3 gives the values $(m, n)$ for the average file size used, as well as conditions on $W_R$ and $\beta^*$. The table also gives the sequences of flight sizes.

NOTE: When the link is saturated, we first calculate the $(m, n)$ as above but we renormalize these values to the known $RTT_0$ so that the new $n \leq \lfloor \beta^* \rfloor$. This renormalization is used for the calculation of the queue contents; it is not necessary for the calculation of $h$ and ON time.

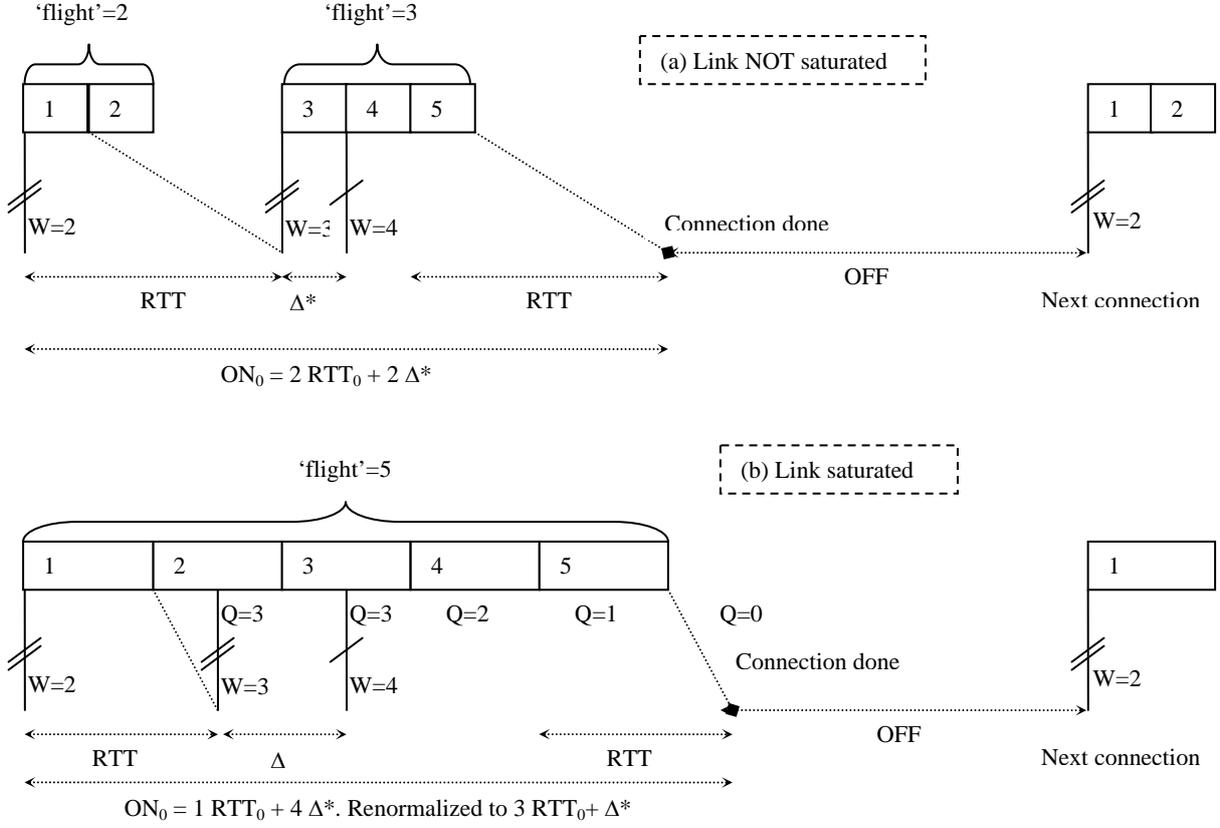

**Figure 6. Illustration of the calculation of $ON_0$ for F = 5 packets.**

**Table 3. Data for the calculation of $ON_0$.**

| F | (m, n), for $ON_0 = mRTT_0 + n\Delta^*$ | 'flight' sizes | Comments |
|---|---|---|---|
| 5 | (2, 2) for $\beta^* > 2$ and $W_R \geq 4$ | (2, 3) | |
| 12 | (3, 5) for $\beta^* > 4$ and $W_R \geq 4$ | (2, 4, 6) | |
| 22 | (4,7) for $\beta^* > 8$ and $W_R \geq 8$ | (2, 4, 8, 8) | |
| 36 | (5, 5) for $\beta^* > 16$ and $W_R \geq 16$ | (2, 4, 8, 16, 6) | |
| | (6, 5) for $\beta^* > 8$ and $W_R = 8$ | (2, 4 , 8, 8, 8, 6) | |
| | (5, 9) for $\beta^* > 8$ and $W_R = 12$ | (2, 4, 8, 12, 10) | |
| | (4,21) for $\beta^* = 8$ and $W_R \geq 16$ | (2, 4, 8, 22) | Saturation on 4th round |
| 80 | (6, 17) for $\beta^* > 32$ and $W_R \geq 32$ | (2, 4, 8, 16, 32, 18) | |
| | (5, 49) for $16 < \beta^* < 32$ and $W_R \geq 16$ | (2, 4, 8, 16, 22) | Saturation on 5th round |
| | (4, 65) for $8 < \beta^* < 16$ and $W_R \geq 9$ | (2, 4, 8, 38) | Saturation on 4th round |
| 120 | (7, 13) for $\beta^* > 44$ and $W_R = 44$ | (2, 4, 8, 16, 32, 44, 14) | |
| | (6, 17) for $32 < \beta^* < 44$ and $W_R = 44$ | (2, 4, 8, 16, 32, 58) | Saturation on 6th round |
| | (6, 17) for $16 < \beta^* < 32$ and $W_R = 44$ | (2, 4, 8, 16, 90) | Saturation on 5th round |
| | (9, 9) for $\beta^* > 20$ and $W_R = 20$ | (2, 4, 8, 16, 20(x 4), 10) | |
| | (16, 9) for $\beta^* > 12$ and $W_R = 12$ | (2, 4, 8, 12 (x 8), 10) | |
| | (17, 1) for $\beta^* > 8$ and $W_R = 8$ | (2, 4, 8 (x 14), 2) | |

For the 'special' settings (100M-1.5M-128k-D300ms-P576B) and $F = 347$ packets, we have $RTT_0 = 341.83ms$, $\beta_2 = 111.3$ packets, $\beta^* = \beta_3 = 9.5$ packets, $\Delta_2 = \Delta = 3.07$ ms, and $\Delta^* = \Delta_3 = 36$ ms. For $W_R > 9$, link $L_3$ is saturated on the $4^{th}$ round starting with packet '15'. We first calculate the flight sizes (2, 4, 8, 333) and $(m, n) = (4, 332)$. We have $h_0 = 120.05$ kbps, $ON_0 = 13319.32ms = 38RTT_0 + 9.16\Delta^*$. We renormalize to $(m, n) = (38, 9)$, which gives $ON_0 = 13313.54ms$. The difference is less than one packet transmission time on $L_3$. For $W_R = 8$, link $L_3$ is not saturated the sequence of flight sizes is (2, 4, 8 [x42], 5) and $(m, n) = (45, 4)$, $h_0 = 103.0$ kbps and $ON_0 = 15526.35ms$.

Note also that with $W_R = 44$, and for file sizes between 54 and 106 packets, a spurious timeout occurs due the way a standard TCP measures the RTT (once per fight at the beginning of the flight). For $F = 54$, a timeout occurs for packet '54', for $F = 106$ it occurs for packet '90'. More specifically, a queue of 19 packets (on link $L_3$) creates a delay of 684 ms and a RTT of 1025 ms. With $W_R = 44$, there are up to 35 packets waiting; that is, up to 1260 seconds of wait time. For a standard TCP in slow start the sequence of timed segments is '1' (W=2), '3' (W=3-4), '7' (W=5-8), '15' (W=9-16), '31' (W=17-32), '63' (W=33-44), '107' (W=44), etc. Between packets '1' and '15' the timed segments see no queue and the RTO remains at 1 second. Packet '31' sees a queue of 23 packets but RTO will be updated only with packet '63'; therefore, a spurious timeout occurs in between. As a consequence, for this setting we compare simulations and calculations for $W_R = 8$ ($L_3$ not saturated) and $W_R = 12$ ($L_3$ slightly saturated with 3 packets waiting).

Once the $ON_0$ time for a single source is determined, we use a simple fluid approximation to calculate $h_0$. Specifically,

$$h_0 = F \times P/ON_0.$$

### 4.3   Calculation of the TCP-Engset $P_j(s)$

We rephrase here the equations of [1].

Let $N$ the number of sources, $n(t)$ the number of sources active (ON) at time $t$. We are interested in $P_j = \mathbb{P}[n(t) = j]$ for $t \to \infty$. The standard theory of BD processes gives $P_j = P_0 \frac{\lambda_0 \lambda_1 \dots \lambda_{j-1}}{\delta_1 \delta_2 \dots \delta_j}$. The $\lambda_j$ and $\delta_j$ are calculated as follows.

The transition rate $j \to j + 1$ is given by

$$\lambda_j = (N - j)/OFF. \tag{9}$$

For a single source, the transition rate OFF to ON is $\lambda = 1/OFF$. When $j$ sources are ON, $N - j$ are OFF. Thus, the total transition rate $\lambda_j = (N - j)\lambda = (N - j)/OFF$.

For the TCP-Engset model, the transition rate $j \to j - 1$ is given by

$$\delta_j = \begin{cases} \frac{j}{ON_0}, & j \leq s = \left\lfloor \frac{C}{h_0} \right\rfloor \\ \frac{C}{F \times P}, & j > s \end{cases}. \tag{10}$$

We first consider the case $j \leq s$. For a single source, the transition rate ON to OFF is $\delta = 1/ON_0$. When $j$ independent sources are ON the total transition rate is $\delta_j = j\delta = j/ON_0$.

For the case $j > s$, we assume perfect and instantaneous capacity sharing between the $j$ sources; that is, the rate per source is $h_j = C/j$. The ON time per source is now $ON_j = \frac{F \times P}{h_j} = j\frac{F \times P}{C}$. The rate of transition per source is $\delta = 1/ON_j$. The total rate for $j$ sources ON is $\delta_j = j\delta = C/FP$.

NOTE: For the calculation of $ON_j$, it is implicitly assumed that the number of active connections does not change widely during the course of the connection, i.e., the ON time.

Finally, we calculate $P_j$ as follows. Let $\pi_0 \equiv 1$, $\pi_j = \pi_{j-1} \times \frac{\lambda_{j-1}}{\delta_j}$ for $j \geq 1$. With $P_0 = 1/\sum_0^N \pi_k$,

$$P_j = \pi_j/P_0. \tag{11}$$

### 4.4   Queue statistics with a single ONOFF connection

We consider here a single connection transferring a file of fixed size $F$ packets. The receiver window is $W_R$. We also consider a capacity mismatch with a link having a packet transmission time $\Delta$ and a BDP $\beta$. In case of simultaneous arrival and departure, priority is given to departure. Finally, the upstream capacity is assumed large so that groups of two packets sent in slow start arrive together. The purpose of this section is to evaluate the 'stress' put on a buffer by TCP slow start. Although the use of slow start may lead to drops (see section 5.4.3), it is less 'damaging' than sending up to the full receiver window of data at once.

When the link is saturated, that is, $W_R > \lfloor\beta\rfloor$, we already know that the long term queue contents $Q = W_R - \lfloor\beta\rfloor + 1$ and $Q_{wait} = W_R - \lfloor\beta\rfloor$, the long term number of packets waiting for transmission.

Let us consider the operation of TCP in 'rounds'. Round '1' starts with the sending of the first two packets of the connection. The queue contents jumps to $Q = 2$. After the transmission of the first packet it decreases to $Q = 1$. Finally, $Q = 0$. In round '2' the receipt of the two acks from the packets sent in round '1' make the sender window increase to $W = 4$, the number of increases is $\nu = 2$, corresponding to the two acks received. The connection proceeds in slow start until the connection is completed or the receiver window is reached.

In Figure 7, we illustrate the 'stress' put on the buffer for $\nu = 4$ and a limited $W_R$.

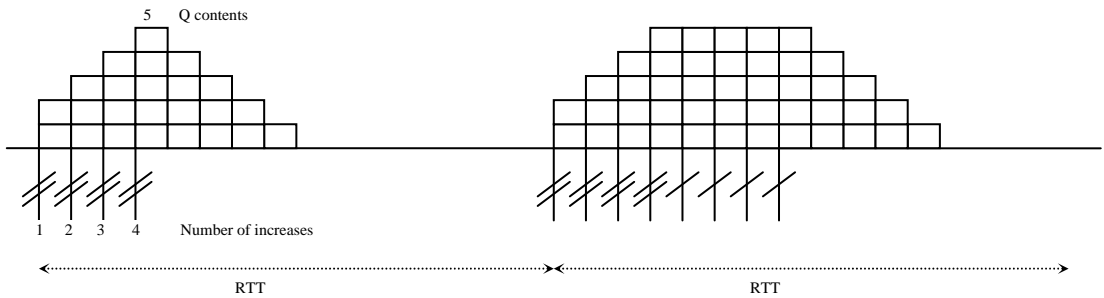

**Figure 7. Illustration of the queue contents with four increases up to the receiver window, link not saturated.**

We now turn to the calculation of the average queue contents using the time average method. The total time is $ON_0 + OFF$ with $ON_0$ calculated as per section 4.2 above. We decompose the calculation of $\bar{Q}$ in contributions per round; that is,

$$\bar{Q}(F) = \frac{\sum_{i=1}^{\#rounds} \tau_i}{ON_0 + OFF}, \text{ where } \tau_i \text{ is the contribution of round } i.$$

For a generic round we have.

$Q_{max} = \nu + 1$ with $\nu$ the number of window increases in the round.

The contribution of the generic round is split in 3 cases.

*1 – Receiver window not reached and link not saturated in the round.*

$\tau = \nu(\nu + 2)\Delta.$

Coming from $\tau = [(\nu + 1) + 1 + 2\sum_2^\nu i]\Delta.$

The equation applies also when the file transfer ends in the round with $F$ even. If $F$ is odd, there is a single last packet and the contribution is $\tau = [(\nu + 1)(\nu + 2) - 1]\Delta.$

*2 – Receiver window reached and link not saturated in the round.*

$\tau = [(\nu + 1)(n + 1) - 1]\Delta$, where $n$ is the number of acks received from the previous round.

NOTE: If $W_R$ is already reached then $\nu = 0$ and $n \leq W_R$. We have $\tau = n\Delta$ and $Q_{max} = 1$, the packet in transmission.

*3 – Link saturated.* The round extends from the first packet sent in the round until the end of the file transfer. We approximate $\tau$ as

$\tau \cong [\nu(\nu + 2) + (F - L_{SS})(\nu + 1)]\Delta,$

where $L_{SS}$ (last in slow start) is the sequence number of the second packet when $W_R$ is reached for the first time. Without delayed acks $L_{SS} = 2W_R - 2$. $F$ is the sequence number of the last packet of the file (our TCP is packet-based and file sizes are in packets). The first term corresponds to the ramps up and down to/from $Q_{max} = \nu + 1$ and the second to the holding time at $Q_{max}$ when $W_R$ is already reached and a leaving packet is replaced by a new one.

EXAMPLE: For the settings (100M1.5M128k-D300ms-P576B-F347) and $W_R > 9$, link L$_3$ is saturated on the 4$^{\text{th}}$ round. Let $W_R = 44$. $L_{SS} = 86$ and $F = 347$. The first three unsaturated rounds contribute $(3 + 8 + 24) = 35\Delta$. For the saturated fourth round we have $\nu = 44 - 8 = 36$ increases, $Q_{max} = 37$. The 4$^{\text{th}}$ round contributes $36 \times 38 + 37(347 - 86) = 11025\Delta$. With $ON_0 = 13319$ ms, $OFF = 5000$ ms, $\Delta = 36$ ms, we have $\bar{Q} = 21.7$ packets (21.2 measured with exponential file sizes), which correspond to 779.9 ms. Thus, the average RTT is increased from $RTT_0 = 341.83$ ms to $RTT = 1121.73$ ms.

With $W_R = 12$, $L_{SS} = 22$, $\nu = 4$, and $\bar{Q} = 3.3$ (2.4 measured) packets corresponding to 118.5 ms only.

# 5  Validation and limitations of the TCP-Engset model

In this section, we verify the validity of the above theoretical considerations as well as their limitations using simulations. We first evaluate the model results – connection rate and utilization – with no packet drops first with exponential file sizes. Second, still with no packet drops, we analyze the effects of file size distribution and receiver window. We then compare theory and simulations with respect to the state probabilities. Finally, we evaluate the model when packet drops occur due to a finite buffer. We also propose a buffer setting rule based on the model.

Throughout this section, the average connection rate is measured in the following three ways.

$$h_{meas} = \frac{1}{N}\sum_1^N \frac{n2send\_tot(i)}{dur\_tot(i)} \cong \frac{\sum_1^N n2send\_tot(i)}{\sum_1^N dur\_tot(i)} = \left(\frac{F \times P}{ON + P/C_1}\right)_{BD},$$

where $n2send\_tot(i)$ is the total of number of packets to send per connection, $dur\_tot(i)$ is the total of connections duration; both for source $i$. $N$ is the number or sources. The first two expressions above give almost the same results with very minor differences owing to the large number of connections completed. The third term is calculated from the results of the BD process and is used as a cross-check.

### 5.1 Exponentially distributed file sizes

The overall settings used in simulations are as follows. Packet size is 1500B. $W_R = 44$ packets. The network settings are (100M2M1G-R50&R100) with $F = (5, 12, 22)$ packets. For these average sizes no link is saturated and no spurious TO occur. The other settings are (100M10M2M-R50&R100) and (100M10M1G-R50&R100) with F=5, 12, 22, 36, 80, 120 packets. For the first setting the egress link is saturated when $F \geq 36$ but no spurious TO occur.

The number of sources is varied to obtain an open-loop traffic demand starting at about 40% up to about 140% by steps of 20%. This corresponds to 6 values of $N$ for each setting.

In Figure 8, we plot the percent error on the calculated connection rate, $h_{(1)}$, $h_{(2)}$, and the average, $h$ versus $\rho_0$. $h_{(1)}$ tends to be larger than measured while $h_{(2)}$ tends to be smaller. Although for some cases $h_{(1)}$ is closer to measurements than $h_{(2)}$ and inversely, as mentioned, we use the average to obtain an overall more accurate result, which is also confirmed by the trend line. Note also that as the load increases, both tend to be accurate as well as equal. The same observation applies to the utilization $\rho$.

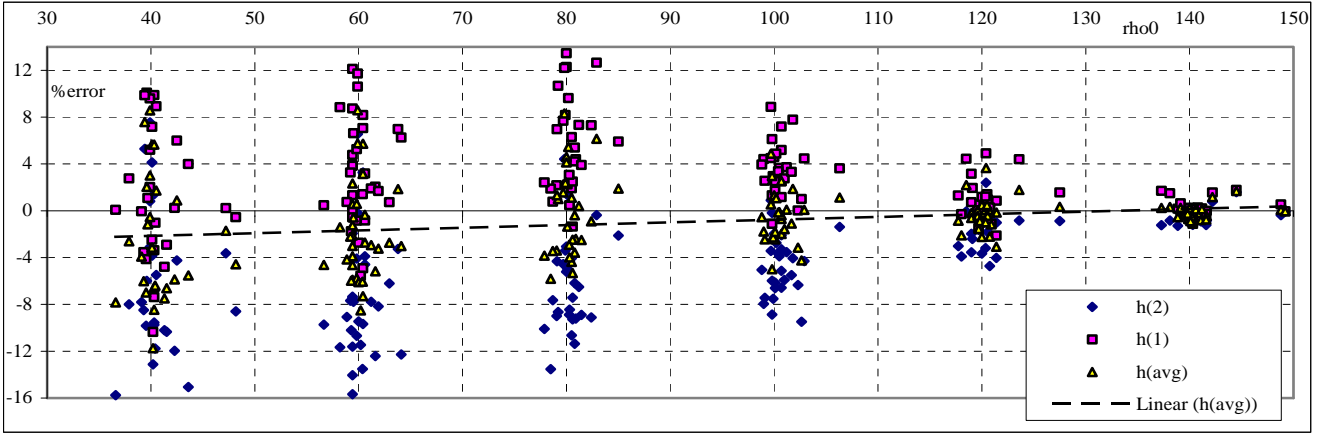

**Figure 8. Percent error of calculated connection rate versus open-loop load.**

Overall, for exponential file sizes, the model is accurate for predicting the connection rate, ON time and link utilization for low to high load with $W_R = 44$.

### 5.2 Effect of file size distribution and receiver window

In this section we analyze the interplay between $W_R$ and file size distribution. The simulation settings are (100M10M1G-R50). The open-loop utilization is 80%, 100%, and 120%. $W_R$ is set so that the calculation of $h$ is not changed with $F$. For example, flight sizes are (2, 4, 8, 8) for $F = 22$ and are unchanged whether $W_R = 44$ or 12. The reference is exponentially distributed file sizes with $W_R = 44$.

In Figure 9 we plot the ratio (in percent) of measured connection rates. Figure 9 left shows that when sizes are exponentially distributed reducing $W_R$ has little influence whatever $F$ and the load. On the contrary, the middle and right plots show that when the size distribution has a longer tail than exponential, $W_R$ has an important effect: reducing $W_R$ increases the average connection rate. The penalty incurred when $W_R = 44$ decreases with increasing $F$.

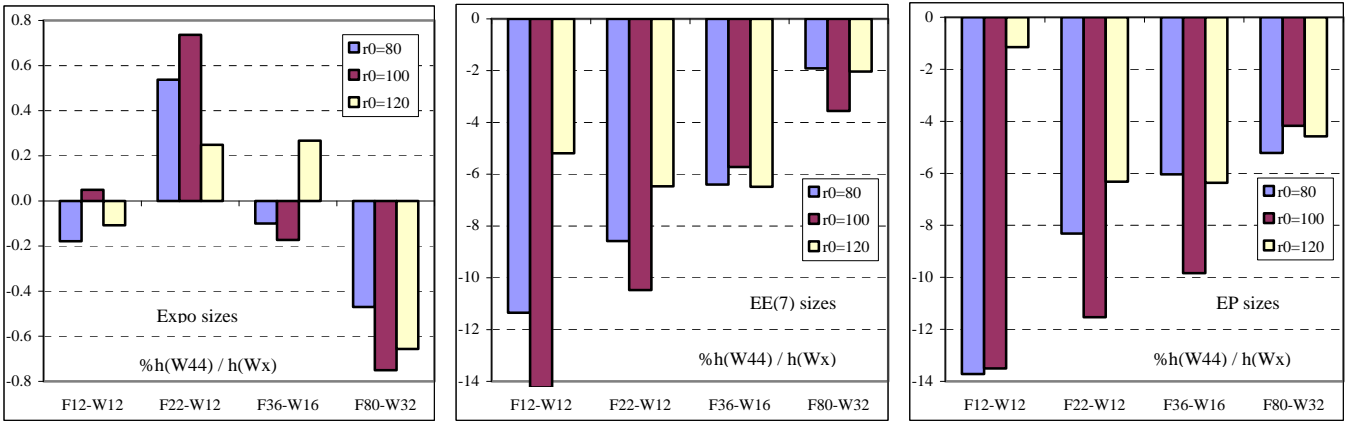

**Figure 9. Percent ratio of measured connection rates per size distribution.**

Using the same simulation data, in Figure 10 we plot the ratio of average connection rate with $W_R = 44$. The reference is the exponential distribution. Again, the penalty decreases with increasing $F$.

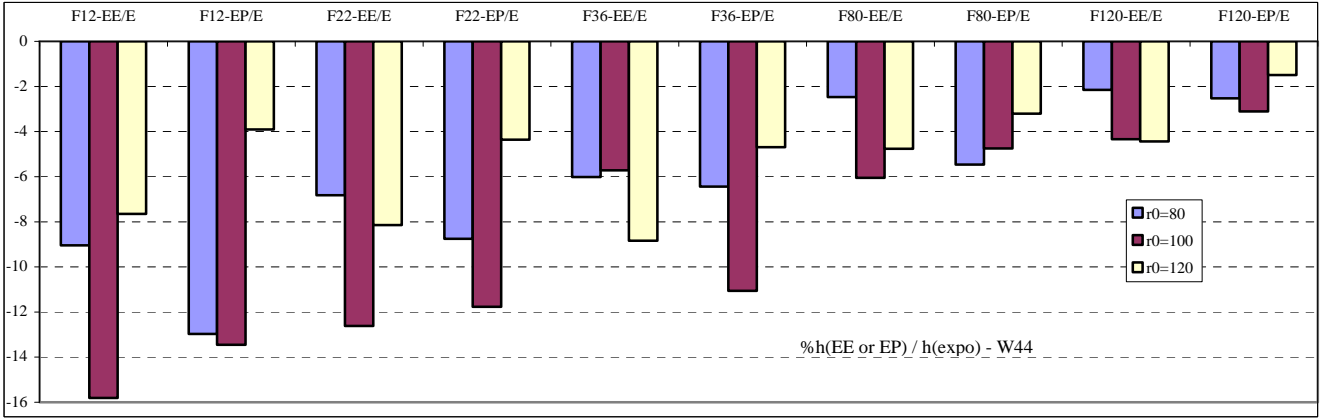

**Figure 10. Percent ratio of connection rate compared to exponential size with $W_R$=44.**

Another perspective is obtained by analyzing the simulation results by size intervals. In Table 4 we give the ratio of ON times for $W_R = 44$ and $W_R = 12$ and ratio of average queue seen per packet for $F = 12$ and $\rho_0 = 100\%$. The effect of long tail distributions and larger $W_R$ is to greatly increase the connection duration for small connections, which are also the most numerous (more than 90% of the total). For files larger than 62 packets, as expected, connections take longer to complete with $W_R = 12$ as indicated by the negative ratio. We have verified that similar results hold also for the (100M10M2M-R50) setting where packets are paced by the egress link, and the (2M10M100M-R50) setting where packets are "naturally" paced by the ingress link. It is thus a generally applicable observation.

We interpret these results by the effect of slow start which results in the increase of the buffer contents and thus increased delay. Recall that in slow start two packets are sent for one leaving the network. With $W_R = 44$ the last packet sent in slow start has sequence number $L_{SS} = 86$ while with $W_R = 12$ it is $L_{SS} = 22$. For $F > L_{SS}$ the effects of larger file sizes are minimal as one packet is sent for one leaving. On the contrary, for $F < L_{SS}$ the size distribution is of importance as files larger than $F$ tend to occupy the buffer and lengthen connections with smaller sizes.

**Table 4. Measured ratios for $W_R$=44 and $W_R$=12.**

| F12-10M1G-R50-N80 ($\rho_0 = 99.7$) | | | | | | | |
|---|---|---|---|---|---|---|---|
| EE (7.4% < 0) | | | | | EP (4.2% < 0) | | |
| f1-f2 | %files | %ON W44/W12 | %Qseen W44/W12 | | %files | %ON W44/W12 | %Qseen W44/W12 |
| 3-6 | 90.19 | 31.1 | 66.0 | | 43.67 | 18.9 | 47.6 |
| 7-14 | 0.89 | 29.6 | 59.9 | | 41.2 | 18.6 | 45.3 |
| 15-30 | 1.55 | 19.1 | 52.0 | | 10.99 | 14.2 | 39.6 |
| 31-62 | 2.34 | -1.6 | 51.1 | | 2.72 | -3.5 | 40.4 |
| 63-128 | 2.73 | -27.9 | 64.9 | | 0.93 | -29.9 | 60.1 |
| 129-256 | 1.79 | -42.0 | 89.4 | | 0.32 | -46.3 | 82.7 |
| 257-512 | 0.48 | -48.2 | 103.7 | | 0.11 | -48.5 | 121.2 |
| 513-1024 | 0.027 | -46.6 | 151.7 | | 0.041 | -49.5 | 141.2 |
| 1025-2048 | | | | | 0.015 | -49.5 | 158.7 |
| 2049-4096 | | | | | 0.005 | -48.5 | 167.2 |
| >4096 | | | | | 0.0026 | -62.2 | 167.9 |

As concerns the TCP-Engset model, its results are accurate provided $W_R$ is adjusted adequately with respect to $F$. The reason is that it does not take into account the relationship between size distribution and slow start. For example, with $F = 12$ and the EE distribution the errors of the model for the three loads of the (100M10M1G-R50) setting are (+11.6%, +19.4%, +7.8%) with $W_R = 44$ but only (+0.2%, +4.5%, +2.5%) with $W_R = 12$. The '+' sign means that the model gives a larger value than measured. The clear advantage of reduced $W_R$ is that most connections complete more quickly thanks to smaller queuing delay.

### 5.3 Sensitivity of $P_j$ to file size distribution and receiver window

We measure the state probabilities, $P_j$, as a time average; that is, as proportion of total time in a given state $j$.

The following Figure 11 shows the comparison between calculations and simulations for the settings indicated in the figure (s is for C/h, b is for BDP). On the left side, the figure shows the state probabilities $P_j$; note the logarithmic scale. On the right side we plot the capacity sharing between ON sources; that is, $\frac{jh_j}{C} = j\left(\frac{P \times F_j}{ON_j}\right)/C$, where $F_j$ and $ON_j$ are the measured values with more than 10 samples. For each case, the number of sources corresponds to an open-loop load of about 100%, which is not a very high load in the present work.

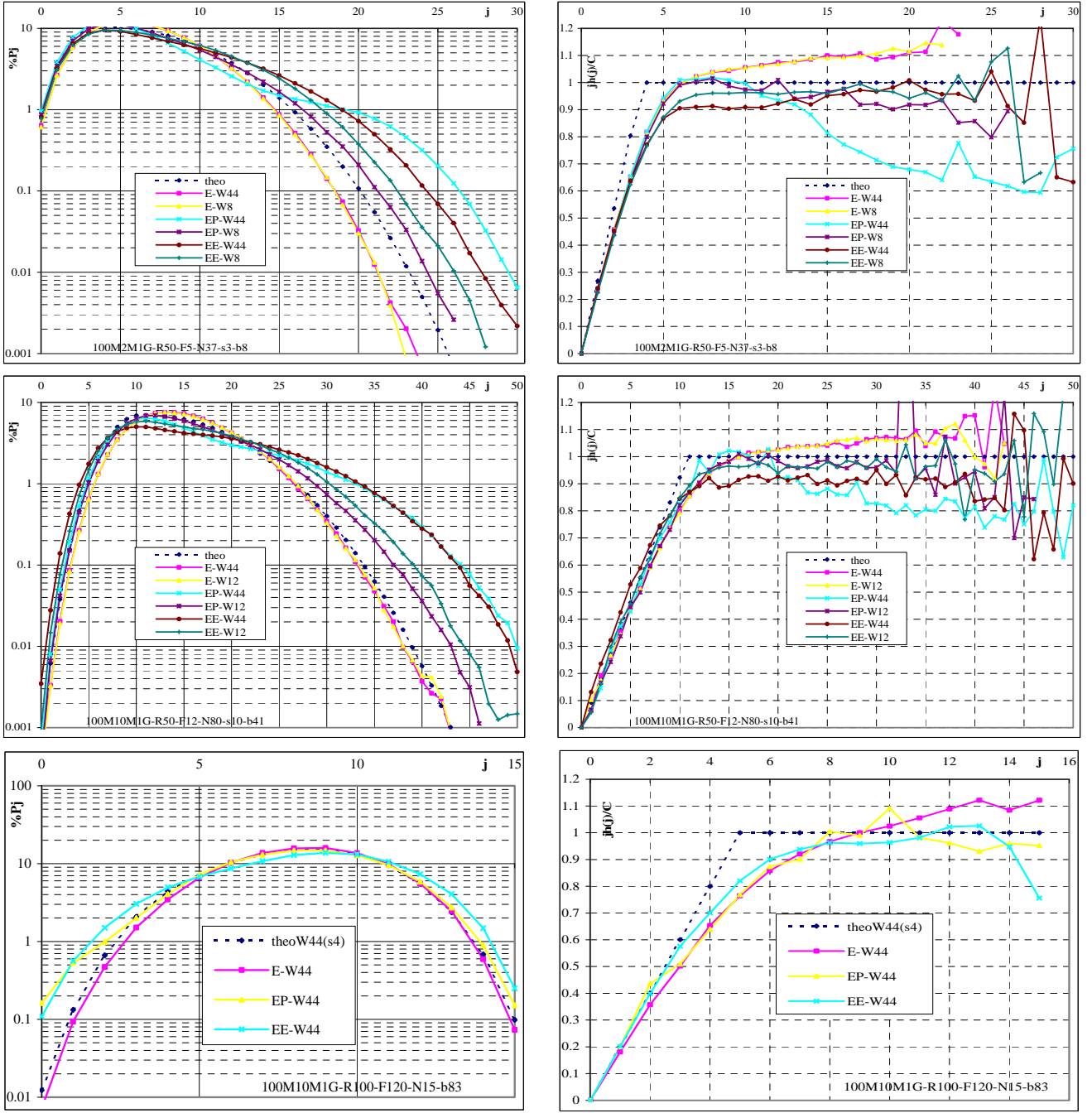

**Figure 11. Comparison between calculations and simulations for different settings. Left, P_j. Right, capacity sharing.**

As expected, the $P_j$'s are not insensitive to the size distribution and $W_R$ for $j > s$. For exponentially distributed sizes, model and simulations agree quite well over the whole range of $P_j$. The reason is that the sizes of files do not go well beyond the average; see the data in Table 1. With longer tail distributions, somewhat larger-than-average sizes appear more frequently, associating this fact with large $W_R$ (relative to the average size) lead to much more queuing occurring (as is measured) and longer connection completion time thus longer holding times in a state $j > s$. Reducing $W_R$ helps lowering the tail of the $P_j$ distribution; however, the interplay between file distribution and $W_R$ remains complex.

Surprisingly, when the average size gets larger ($F = 120$ or $F = 347$, not shown here) the insensitivity seems weaker for $j \leq s$. The reason is neither queuing nor the lack of samples. We cannot explain this difference.

The plots on the right side of Figure 11 show that while capacity sharing occurs, it is not as steeply marked as the model predicts.

As seen in the previous section, the model holds good for exponential size distribution. For longer tail distributions, it works well when $W_R$ is adjusted with respect to $F$.

### 5.4 Performance with limited buffer

In this section, we first propose and evaluate a buffer setting rule based on the TCP-Engset model. We then turn to the analysis of the effects of packet drops. Finally, we analyze the effects of varying the buffer size.

#### 5.4.1 Buffer setting rule

Consider a single multiplexing link of capacity $C$ with a round trip time $RTT$. One rationale for choosing the buffer size $B$ for the link is to avoid spurious TCP timeouts. With an RTO of 1000 ms, allowing a maximum queuing delay of 600 ms leaves 400 ms for the $RTT$. We call this setting the 600ms-rule. A second rationale is to set the buffer to the BDP. This setting considers long connections in congestion avoidance with synchronized windows. The delay is usually set to 300 ms; we call this setting the BDP-rule. This rule is analyzed in [1] as well as in a number of other publications.

We propose a buffer setting rule based on the TCP-Engset model developed above. Set $L = 1\%$ and $W_R = 44$. For $N$ such that $\rho_0 \cong 100\%$, and $RTT = 50$ and $RTT = 300$ ms, compute $B = \eta \text{Ln}(\rho)$, with $\eta$ and $\rho$ given by the model. For each $F$ the calculation gives a different $B$. We round up the results obtained for $F = 5, 12, 22, 36, 80, 120, 500,$ and $1000$ packets. There is no clear cut tendency, it is therefore recommended to vary the parameters for the calculation. For a 10 Mbps link the model gives $B = 202$ for $F = 500$. For a 50 Mbps it gives $B = 417$ also for $F = 500$.

In Table 5 we give the buffer size in packets and the maximum queuing delay for different link capacities and rules. The TCP-Engset-rule gives smaller values than the other rules especially for the 50 Mbps link.

**Table 5. Buffer size in packets for RTT=300 ms and P=1500B.**

| C (Mbps) | 600ms-rule | BDP-rule | TCP-Engset-rule |
|---|---|---|---|
| 2 | 100 (600 ms) | 50 (300 ms) | |
| 10 | 500 (600 ms) | 250 (300 ms) | 200 (240 ms) |
| 50 | 2500 (600 ms) | 1250 (300 ms) | 420 (100 ms) |

#### 5.4.2 Queue behavior

In Figure 12 we plot the theoretical and measured $\mathbb{P}(Q > x)$ for the settings indicated in the figure; both are for $\rho_0 = 100\%$. For the theoretical (T) results we give the pair $(\rho, \eta)$, and the calculated $L$ and $h$. The measured values indicated are $L$, $\rho$ (noted r), and $h$.

First, the queue contents greatly depend on size distribution and receiver window. As in the previous sections, we interpret this observation by the effects of slow start (two packets sent for one leaving) on the buffer and the relationship between $F$ and $L_{SS}$. Second, the approach proposed here for $\mathbb{P}(Q > x)$ works fairly well for exponentially distributed sizes and for the EP distribution with $W_R = 12$. Note also that for the E distribution the measurements are well below the model in the tail. Third, as concerns the loss rate, the model provides an upper bound for the E distribution with $W_R = 44$ and the EE and EP distributions with $W_R = 12$. Finally, small loss rates do not affect the accuracy of the predictions of $\rho$ and $h$.

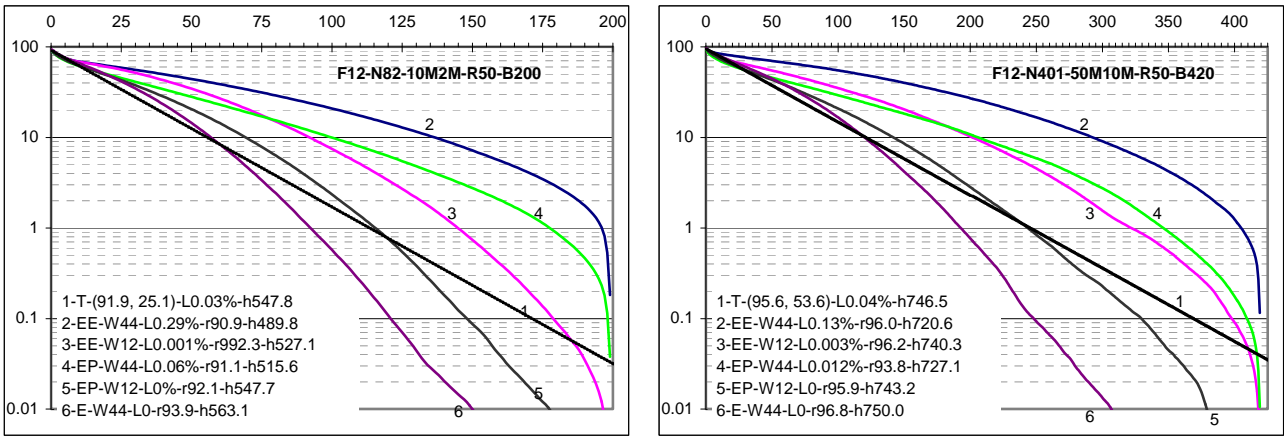

**Figure 12. Theoretical vs. measured P(Q>x) for limited buffer.**

The next question is: "what happens when the number of sources increases with the same buffer setting?"

In Table 6 we give the comparison between theory and measurements for the setting indicated. The first observation is that connection rates tend to equalize with increasing load. Further, they tend to the theoretical value. We have reproduced the settings of [1] with the same outcomes. The second observation is that the model greatly overestimates the loss rates above 1%.

**Table 6. Comparison of results for different loads.**

| 100M10M2M-R50-$W_R$44-B200 | | | | |
|---|---|---|---|---|
| | N82 $\rho_0 = 100.1\%$ | N98 $\rho_0 = 119.6\%$ | N115 $\rho_0 = 140.3\%$ | N134 $\rho_0 = 163.5\%$ |
| Theory-h (kbps) Theory-L (%) | 547.8 0.03 | 344.6 4.91 | 219.5 22.0 | 154.9 38.3 |
| EE(7)-h (kbps) EE(7)-L (%) | 489.8 0.29 | 328.2 1.04 | 219.2 2.66 | 155.9 4.80 |
| E-h (kbps) E-L (%) | 563.1 0 ($Q_{max}$175) | 347.1 0.004 (399/9.6M) | 219.5 0.41 | 154.7 2.55 |

To conclude this topic on the buffer contents, in Figure 13 we plot the queuing behavior for $\rho_0 = 80.5\%$ (moderate load) and $\rho_0 = 119.6\%$ (high load) for $B = 200$ packets. For the moderate load, the model fairly follows measurements for the exponential distribution. It greatly underestimates the buffer contents for the EE distribution, even with $W_R = 12$. For the high load, even though the difference between model and measurements are reduced, the 'bulge' above the theoretical curve is not taken into account.

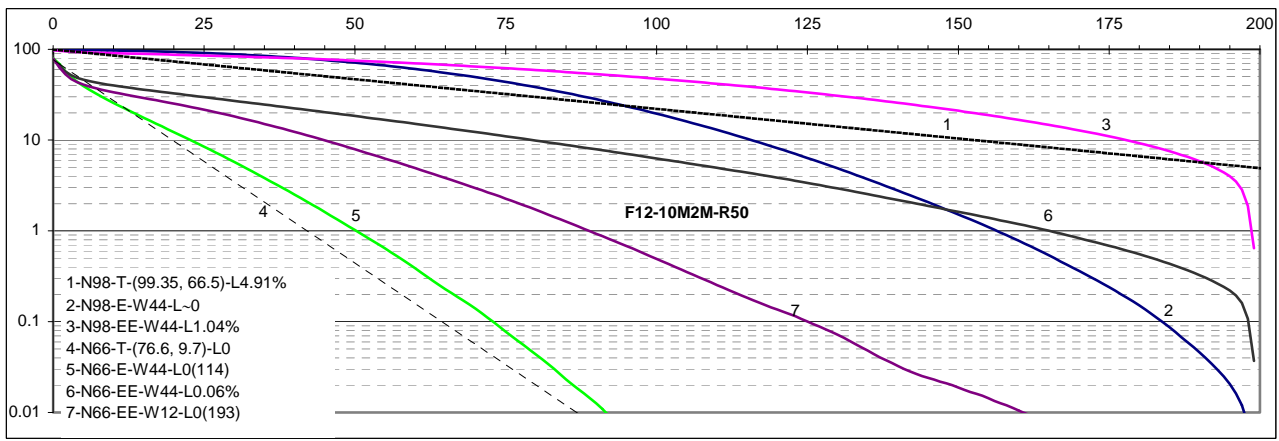

**Figure 13. Queuing behavior for a moderate load and a high load.**

The same plots for the (2M10M100M-R50) setting, that is, when packets are paced by the ingress link, are almost indiscernible from those of Figure 13. The observations are thus of general applicability.

### 5.4.3   Analysis of losses

We now turn to the analysis of losses. Here, packets are only dropped due to buffer overflow; there are no random drops. In Figure 14 we plot the proportion of drops per sequence number (recall that each connection starts with packet '1') for a high load and high losses setting indicated in the figure. When the connection is in slow start the second packet sent gets more dropped than the first by a large factor. This is due to the capacity mismatch between the 100 Mbps ingress link and the 10 Mbps multiplexing link. The observation does not hold when packets are 'naturally' paced by an ingress link of 2 Mbps; although other measurements results are similar (in particular, loss rate and connection rate). The effect of lower $W_R$ is to reduce the loss rate from 6.3% to 4.6%. Note also that packets are more uniformly dropped after the last packet sent in slow start (packet '86' for $W_R = 44$ and '22' for $W_R = 12$). We do not use this observation here.

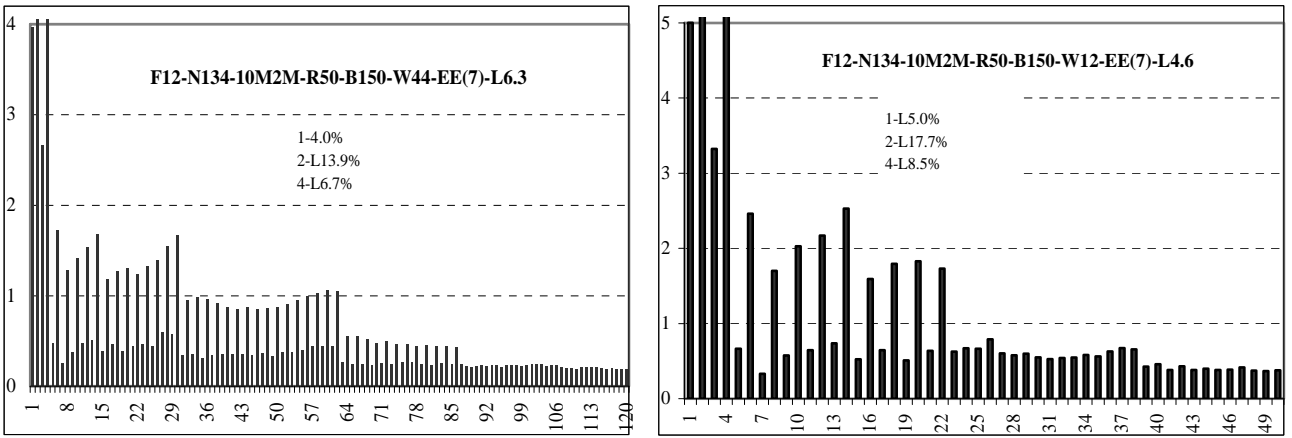

**Figure 14. Proportion of dropped packets per sequence number.**

Packet drops have two consequences on TCP. Drops are recovered either by TO or FR (a TD followed by a successful fast recovery). With one drop only, a TO is unavoidable if packets '1' or '2' are dropped, we call this event TO beginning (TObeg). At the end of the connection, if either one of the three last packets are dropped a TO follows; the event is called 'TOend'. If any of the retransmissions after TO, or during the fast recovery, is dropped a timeout follows. A TO after a TO is called TOTO and it occurs at $W = 1$. A TO during fast recovery is called TDTO. With the 'impatient' version of TCP NewReno this event only occurs if either one of the retransmissions is dropped. In Table 7 we give the drop analysis for the high load/loss setting of Figure 14 left.

We start with TDFR events. From the results of Table 7 we see that FR occurs most commonly; only 3% of TDs are followed by a TO. Packets (including retransmissions) continue to be sent at a rate similar to the overall send rate. Thus, connections are not stalled by TDFR events; as it should. In Figure 15 we plot the distributions of the number of retransmissions (left) and the number of packets sent on FR (right) for $W_R = 44$ and $W_R = 12$. Our TCP resets the timer on each retransmission; this is to avoid spurious TOs during fast recovery. Our TCP does not limit the burst of packets to send; Figure 15 right shows that most significant bursts are of 2 (82.8%) or 3 (6.1%) packets for $W_R = 44$ and up to 7 (11.5%) for $W_R = 12$. Limiting the bursts to 2 packets, as in slow start, does not improve performance in a significant way.

**Table 7. Drop analysis for a high load/loss setting.**

| F12-10M2M-R50-EE(7)-W44-B150-L6.3% | |
|---|---|
| %packet drops | 6.3 |
| | |
| %TO per dropped packet | 26.5 |
| %TOTO per TO | 9.9 |
| %TDTO per TO | 2.4 |
| %TObeg per TO | 23.0 |
| % TOend per TO | 61.4 |
| % TOother per TO | 3.4 |
| Avg time TO per TO (ms) | 1200.5 |
| Avg time TO per connection (ms) | 259.0 |
| | |
| %TD per dropped packet | 20.8 |
| %FR per TD | 97.0 |
| Avg nb pkts sent TDFR per FR | 11.1 |
| Avg nb pkts sent on FR per FR | 2.3 |
| Avg time TDFR per FR (ms) | 614.3 |
| Avg time TDFR per connection (ms) | 100.8 |
| Avg rate TDFR (kbps) per FR | 216.8 |
| | |
| Avg send rate (kbps) | 197.0 |
| Avg connection rate, $h$ (kbps) | 154.5 (154.9 theory) |
| Avg connection ON time (ms) | 927.1 (929.6 theory) |
| Avg file size (pkts) | 11.94 (12 theory) |
| Avg RTT measured by TCP (ms) | 196.4 (122 pkts in queue) |

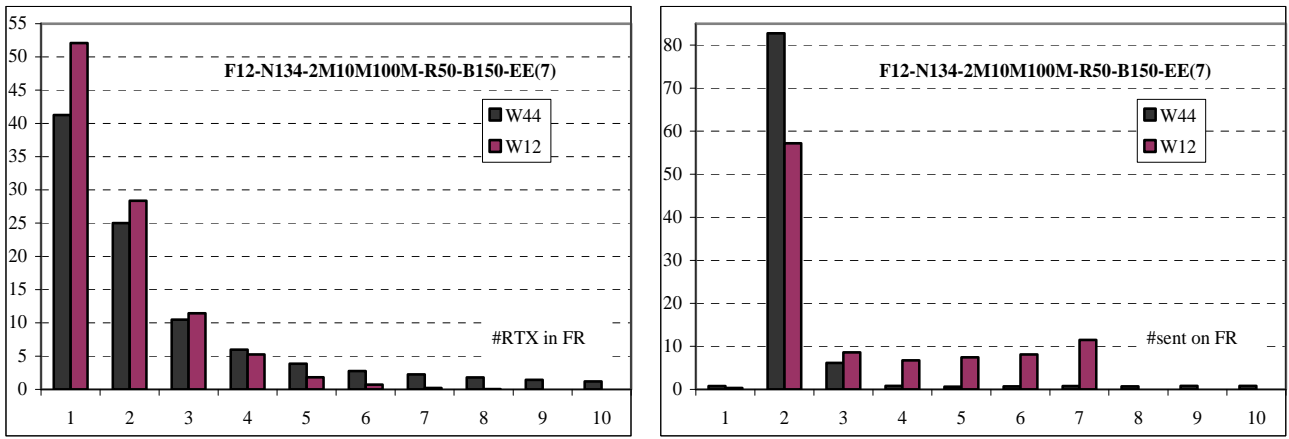

**Figure 15. Left, number of retransmissions in TDFR. Right, number sent on FR.**

We now turn to TO events. The occurrence of timeouts has more consequences on TCP performance as a connection experiencing a TO gets stalled for about 1 second and possibly more. We focus on TOs in slow start as they constitute the vast majority of these events. In our TCP, a connection is in slow start when $W \leq S$. In Figure 16 we plot the proportion of TOs versus the window at which they occur for the high loss setting for the EE (left) and E (right) size distributions; both for $W_R = 44$. A TO at $W = 1$ indicates a TOTO. A TO at $W = 2$ means that packet '1' is dropped. A TO at $W = 3$ means that packet '1' is not dropped but packet '2' is; whatever happens to packets '3' or '4'. A TO at $W = 4$ means that packet '3' is dropped and, at least, one more drop occurs on packets '4', '5', or '6'. A TO at $W = 5$ requires three drops starting with packet '4', etc. The proportion of TOs depends on the size distribution; for example, more than 90% of files are between 3 and 6 packets for the EE(7) distribution. Clearly, the performance penalty is high for small connections that should complete in a few RTTs and experience one or more TOs.

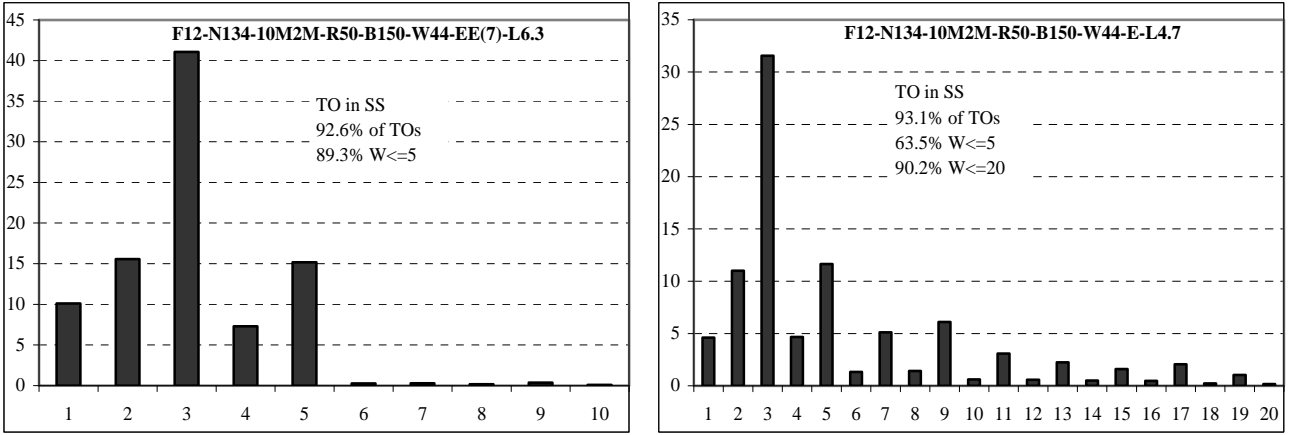

**Figure 16. Proportion of TO in slow start vs. window. Left, EE(7) distribution. Right, E distribution.**

We turn to the loss analysis per file size. In Table 8 we show the measured statistics per file size interval for the EE(7) distribution. The vast majority (90.3%) of sizes are between 3 and 6 packets. Larger sizes are more evenly spread. The average queue seen per packet is also fairly constant and about one packet less than the overall time average. The 6[th] column gives the proportion of TOs per packet dropped. For the 3-6 packets connections a single drop leads almost inevitably to a TO as a TDFR sequence cannot occur. That proportion decreases with increasing file size. The 7[th] column gives the proportion of drops recovered by a TDFR (number of retransmissions over number of drops). The dependence on file size is the inverse of that of TOs per drops.

**Table 8. Statistics per size interval for a high losses setting.**

| N134-10M2M-R50-EE(7)-W44-B150. Measured: $F$=11.88 pkts, $h$=155.5 kbps, avgQ=126.7 pkts | | | | | | |
|---|---|---|---|---|---|---|
| f1-f2 | %prop. | $<f>$ | $h$ (kbps) | $<Q_{seen}>$ per pkt | %TO per drop | %rtx FR per drop |
| 3-6 | 90.27 | 3.4 | 93.4 | 125.3 | 83.9 | 0.0 |
| 7-14 | 0.87 | 10.5 | 161.6 | 124.3 | 43.4 | 28.9 |
| 15-30 | 1.55 | 22.3 | 222.8 | 125.8 | 21.4 | 61.7 |
| 31-62 | 2.31 | 45.5 | 292.3 | 127.4 | 11.3 | 78.2 |
| 63-128 | 2.73 | 91.1 | 349.5 | 128.4 | 7.5 | 82.3 |
| 129-256 | 1.77 | 176.3 | 398.3 | 128.7 | 6.2 | 84.2 |
| 257-512 | 0.48 | 326.0 | 418.1 | 128.6 | 5.6 | 86.4 |
| 513-1024 | 0.023 | 590.4 | 396.2 | 128.5 | 4.4 | 89.7 |

### 5.4.4 Effect of varying the buffer size

We now turn to the effects of varying the buffer size. The difference between small and large buffers can be seen in Table 9. For each file size interval we give the connection rate $h_2$ (see below) and, in parenthesis, the proportion of the ON time spent/stalled in TO (*time_TO – time_previousSent*) for different buffer sizes. The measured values are obtained as follows ($N_c$ is the number of connections).

$$ON = \frac{\sum_{\text{per connection}} duration}{N_c}.$$

$$h = \frac{\sum_{\text{per connection}} n2send}{\sum_{\text{per connection}} duration} = \frac{n2send\_tot}{duration\_tot} = \frac{F}{ON} \neq h_2 = \frac{1}{N_c}\sum_{\text{per connection}} \left(\frac{n2send}{duration}\right).$$

As concerns the connection rate, the results show that small buffers favor small connections and disfavor large ones while large buffers disfavor small connections and favor large ones. However, for a small buffer, the proportion of time where connections are stalled waiting for a TO is large, whatever the file size. Note also that the definition $h_2$ of the average connection rate gives a larger value than the definition $h$. The actual values differ depending on the file size distribution.

**Table 9. Connection rate $h_2$ (kbps) and proportion of TO in ON time per size interval for different buffer sizes.**

| F12- N134-10M2M-R50-W_R44. $h_2$ in kbps (%TO time / ON time) | | | | | |
|---|---|---|---|---|---|
| | EE(7) | | | EP | | |
| f1-f2 (pkts) | B50 | B150 | B750 | B50 | B150 | B750 |
| | L=9.5% | L=6.3% | L=0.001% | L=8.8% | L=4.75% | L=0 ($Q_m$587) |
| 3-6 | 160.9 (77.2) | 92.8 (38.0) | 62.0 | 205.3 (74.1) | 120.5 (32.5) | 93.3 |
| 7-14 | 223.1 (73.7) | 157.9 (37.8) | 121.8 | 253.8 (66.9) | 160.5 (28.1) | 131.2 |
| 15-30 | 290.1 (60.9) | 221.3 (28.2) | 191.8 | 328.0 (56.7) | 220.8 (21.1) | 192.7 |
| 31-62 | 348.6 (52.6) | 287.5 (20.3) | 304.1 | 425.6 (45.6) | 317.7 (14.5) | 323.0 |
| 63-128 | 386.4 (46.1) | 342.3 (15.5) | 471.5 | 478.7 (38.4) | 401.8 (10.8) | 521.6 |
| 129-256 | 400.5 (42.0) | 386.5 (12.1) | 672.8 | 525.4 (32.3) | 464.9 (7.3) | 757.9 |
| 257-512 | 406.9 (38.1) | 406.5 (9.3) | 835.1 | 549.8 (27.5) | 516.1 (5.1) | 975.9 |
| 513-1024 | 390.4 (36.5) | 432.4 (6.3) | 959.9 | 560.5 (23.5) | 510.9 (4.8) | 1128.1 |
| 1025-2048 | | | | 556.6 (22.3) | 522.8 (3.2) | 1216.4 |
| 2049-4096 | | | | 540.2 (22.3) | 518.1 (1.7) | 1270.7 |
| > 4096 | | | | 532.6 (22.5) | 642.9 (1.8) | 1308.9 |

We conclude this section with a surprising observation. In Table 10, for a high load setting, we show the connection duration (ON time), connection rate ($h$), and loss rate ($L$) for different buffer sizes and EE(7) and E size distributions. The surprising observation is that, as concerns the averages, there seems to be compensation between delay and losses to give the same result. For large buffers, connections take a long time to complete because of the delay. For small buffers, the recovery of packet drops lengthens connections. The theoretical values also given show that the model predicts fairly well both the connection rate and duration independently of buffer size and distribution. However, repeating the simulations with $N = 82$ ($\rho_0 = 100.1\%$) show that it is only due to the very high load.

**Table 10. Measured averages for different buffer sizes.**

| F12-N134-10M2M-R50-W44. Theory: $h$=154.9 kbps, ON=929.6 ms, $\rho_0 = 163.5\%$ | | | | | | | | | |
|---|---|---|---|---|---|---|---|---|---|
| | | B50 | B75 | B100 | B150 | B200 | B300 | B500 | B750 |
| | h (kbps) | 154.6 | 154.9 | 155.1 | 155.5 | 155.9 | 155.9 | 156.7 | 156.0 |
| EE(7) | ON (ms) | 924.3 | 921.1 | 922.6 | 917.3 | 914.5 | 917.5 | 910.6 | 916.3 |
| | L % | 9.5 | 8.6 | 7.8 | 6.3 | 4.8 | 2.3 | 0.16 | 0.001 |
| | h (kbps) | 153.0 | 153.6 | 153.7 | 154.2 | 154.7 | 155.4 | 155.1 | 154.8 |
| E | ON (ms) | 939.4 | 937.6 | 937.2 | 933.8 | 929.7 | 926.0 | 927.6 | 930.7 |
| | L % | 9.3 | 8.1 | 7.0 | 4.7 | 2.55 | 0.18 | 0 | 0 |
| | (theory) | (78.7) | (69.8) | (61.9) | (48.7) | (38.3) | (23.7) | (9.1) | (2.7) |

As a final note in this section, a simulation with $N = 134$ takes about 30 minutes to complete while with $N = 401$ it takes about 1h30mn. It is therefore interesting to have a model to quickly compute quantities of interest even though a simulation can supply results that cannot be calculated.

# 6   Summary and Conclusion

In the present work, we have revisited the TCP-Engset model proposed in [1]. The model deals with the performance obtained with a limited number of short connections sharing a multiplexing link. Here, we have considered homogeneous sources only.

We have taken into account the effects of slow start and limited receiver window. We have proposed an alternative way of calculating the average connection rate. Extensive simulations show that taking the average of the two theoretical results give more accurate results than each individual ways. We have proposed an approach for the determination of the queuing behavior. This approach has led us to propose a buffer setting rule. We have validated and determined the limits of the revisited model with extensive simulations.

The approach developed in the present report is summarized and commented in Table 11.

Our conclusions are as follows.
-   All of connection rate, ON time, link utilization, queue contents, state probabilities, and loss rate are sensitive to file size distribution and receiver window.
-   However, this sensitivity tends to vanish with increasing load; whether packets are dropped (here, only due to buffer overflow) or not.
-   Under a wide range of loads, the revisited model predicts fairly accurately the average connection rate, duration, and link utilization for exponentially distributed file sizes. For longer tail distributions, the model remains accurate provided the receiver window is adjusted with respect to the average file size. These distributions lead to reduced results with respect to exponential size and the model; the penalty is up to 15-20%. Further, the most numerous small connections are penalized by the presence of long connections with large receiver window.
-   We have interpreted these observations by the relationship between $F$, the average file size, and $L_{SS}$, the last packet sent in slow start when the receiver window is reached for the first time.
-   As concerns the queuing behavior, the model is fairly accurate under the same conditions (distribution and receiver window) as in the previous point.
-   As concerns the loss rate, the model *cannot* be used to predict loss rates larger than about 1%. The model greatly overestimates the loss rate above that threshold. Further, it does not take into account retransmissions.
-   The buffer sizing rule we have proposed appears to work well under the condition mentioned in the previous point, that is, when loss rates are below 1%.

Clearly, new ideas and more work are required to take into account the influence of file size distribution and receiver window. The same applies for the accurate prediction of the loss rate for TCP. Finally, the case of heterogeneous sources has not been dealt with and is left for future modeling work.

**Table 11. Commented summary of the TCP Engset revisited approach.**

| Steps of the approach for homogeneous sources. | Comments |
|---|---|
| **Calculation of the single connection parameters.**<br><br>Connection ON time. $ON_0 = mRTT_0 + n\Delta^*$, with $\Delta^* = \max{(\Delta_i)}$.<br>Connection rate. $h_0 = \frac{F \times P}{ON_0}$<br>Multiplexing link s-parameter. $s = \left\lfloor \frac{C}{h_0} \right\rfloor$.<br>Open loop link load with $N$ sources.<br>$\rho_0 = \langle n \rangle \times h_0 / C = \frac{ON_0}{ON_0 + OFF} \times N \times h_0 / C$. | $RTT_0$: time between the sending of a single packet of size $P$ and the receipt of its ack, in an empty path.<br><br>$\Delta_i = P/C_i$, $\beta_i = RTT/\Delta_i$.<br><br>Allow $N$ such that both $\rho_0 < 1$ and $\rho_0 > 1$ are possible. |
| **Calculation of the state probabilities $P_j$.**<br><br>$\pi_0 \equiv 1$, $\pi_j = \pi_{j-1} \times \frac{\lambda_{j-1}}{\delta_j}$, with<br>$\lambda_j = (N - j)/OFF$ and<br>$\delta_j = \begin{cases} \frac{j}{ON_0}, & j \le s = \left\lfloor \frac{C}{h_0} \right\rfloor \\ \frac{C}{F \times P}, & j > s \end{cases}$<br><br>$P_j = \pi_j / P_0$ with $P_0 = 1/\sum_0^N \pi_k$. | |
| **Average connection rate, ON time, and link utilization.**<br><br>Let $P_{OL} = \sum_{s+1}^N P_k$, $n_{UL} = \sum_1^s kP_k$, $n_{OL} = \sum_{s+1}^N kP_k$, and $\langle n \rangle = \sum_1^N kP_k = a \times N$.<br><br>From [1], Throughput = $T = h_0 \times n_{UL} + C \times P_{OL}$.<br><br>$\quad h_{(1)} = T/\langle n \rangle$ and $\rho_{(1)} = T/C \le 1$.<br><br>Alternative way.<br>$\quad v_{OL} = \frac{n_{OL}}{P_{OL}} = \frac{\sum_{s+1}^N kP_k}{\sum_{s+1}^N P_k}$, $h_{(2)} = C/v_{OL}$, and $ON = \frac{F \times P}{h_{(2)}}$.<br><br>$\quad \rho_{(2)} = \langle n \rangle \times h_{(2)} / C = \frac{ON}{ON + OFF} \times N \times h_{(2)} / C \le 1$<br><br>Overall, $h = \frac{h_{(1)} + h_{(2)}}{2}$, $ON = \frac{F \times P}{h}$, and $\rho = \frac{\rho_{(1)} + \rho_{(2)}}{2}$. | $h_{(1)}$ and $\rho_{(1)}$ tend to be slightly larger than measured while $h_{(2)}$ and $\rho_{(2)}$ tend to be smaller. The average is more accurate.<br><br>It turns out that $\rho \le 1$ is the link utilization whether drops occur or not. |
| **Queue behavior and link buffer setting, TCP Engset-rule.**<br><br>Closed loop RTT. $RTT = \frac{ON - n\Delta^*}{m}$.<br><br>Average queue contents. $Q = \frac{RTT - RTT_0}{\Delta}$.<br><br>Assume exponential queue contents decay with factor $\eta = Q/\rho$.<br><br>$\quad G(x) = \mathbb{P}(Q > x) = \rho e^{-(x/\eta)}$.<br><br>$\quad$ Loss rate $L = \rho e^{-(B/\eta)}$, so that<br><br>$\quad B = \eta \text{Ln}(\rho/L)$, with $L \cong 1\%$. | When a link on the path can be saturated, renormalize $ON_0$ to $RTT_0$.<br><br>$Q$ and $\Delta$ refer to the multiplexing link.<br><br>**Rule:** Set $N$ such that $\rho_0 \cong 100\%$. Set $W_R = 44$, calculate $B$ for different $F$ up to $F=1000$ and $RTT_0 = 50$ ms and 300 ms. Round up the largest $B$ obtained. For a 10 Mbps link, $B = 200$. For a 50 Mbps link, $B = 420$ packets.<br><br>The queue contents, $G(x)$, is sensitive to size distribution and receiver window.<br><br>The equation for $L$ cannot be used to predict loss rates larger than about 1%. |